\documentclass[a4paper,fleqn]{cas-dc}

\usepackage{bm}
\usepackage{stfloats}
\usepackage{subfigure}
\usepackage{multirow}
\usepackage{diagbox}
\usepackage{enumerate}
\usepackage{threeparttable}
\usepackage{multirow}
\usepackage{booktabs}
\usepackage{color}
\usepackage{url}
\usepackage{makecell}
\usepackage{caption}
\usepackage[authoryear]{natbib}

\def\tsc#1{\csdef{#1}{\textsc{\lowercase{#1}}\xspace}}
\tsc{WGM}
\tsc{QE}
\tsc{EP}
\tsc{PMS}
\tsc{BEC}
\tsc{DE}

\begin{document}
\captionsetup[figure]{labelfont={bf},name={Fig.},labelsep=period}
\let\WriteBookmarks\relax
\def\floatpagepagefraction{1}
\def\textpagefraction{.001}
\shorttitle{Journal of Network and Computer Applications} 
\shortauthors{J. Wu et~al.}
\captionsetup[table]{
  labelsep=newline,
  singlelinecheck=false,
}
\title [mode = title]{Analysis of Cryptocurrency Transactions from a Network Perspective: An Overview}                      



\author[1]{Jiajing Wu}
\ead{wujiajing@mail.sysu.edu.cn}


\address[1]{School of Computer Science and Engineering, Sun Yat-sen University, Guangzhou 510006, China}

\author[1]{Jieli Liu}
\ead{liujli7@mail2.sysu.edu.cn}

\author[2]{Yijing Zhao}
\ead{zhaoyj53@mail2.sysu.edu.cn}

\author[1]{Zibin Zheng}
\cormark[1]
\ead{zhzibin@mail.sysu.edu.cn}

\address[2]{School of Electronics and Communication Engineering, Sun Yat-sen University, Guangzhou 510006, China}

\cortext[cor1]{Corresponding author}

\begin{abstract}
As one of the most important and famous applications of blockchain technology, cryptocurrency has attracted extensive attention recently. Empowered by blockchain technology, all the transaction records of cryptocurrencies are irreversible and recorded in blocks. These transaction records containing rich information and complete traces of financial activities are publicly accessible, thus providing researchers with unprecedented opportunities for data mining and knowledge discovery in this area. 
Networks are a general language for describing interacting systems in the real world, and a considerable part of existing work on cryptocurrency transactions is studied from a network perspective.
This survey aims to analyze and summarize the existing literature on analyzing and understanding cryptocurrency transactions from a network perspective. Aiming to provide a systematic guideline for researchers and engineers, we present the background information of cryptocurrency transaction network analysis and review existing research in terms of three aspects, i.e., network modeling, network profiling, and network-based detection. For each aspect, we introduce the research issues, summarize the methods, and discuss the results and findings given in the literature. Furthermore, we present the main challenges and several future directions in this area.
\end{abstract}




\maketitle
\section{Introduction}
Recent years have witnessed the emergence of a new type of tradable asset called cryptocurrencies. As the fundamental technology underlying cryptocurrency, blockchain provides a distributed and decentralized environment for transactions of the emerging cryptocurrencies including Bitcoin.
Along with the rapid development of blockchain technology, these blockchain-based cryptocurrencies have also gained increasing popularity and attention in the past decade. As of the second quarter of 2020, more than 7,000 cryptocurrencies are actively traded and their total market cap has exceeded 300 billion U.S. dollars.

By employing peer-to-peer (P2P) transmission, consensus algorithms, and incentive mechanisms, the issuance and transactions of cryptocurrencies can be performed without a centralized authority.
Empowered by blockchain technology, all the transaction records of cryptocurrencies are irreversible and recorded in the blocks, which are linked in chronological order.

Due to the open and transparent nature of blockchain, {\color{black}cryptocurrency} transaction records containing rich information and complete traces of financial activities are publicly accessible, thus providing researchers with unprecedented opportunities for data mining in this area. The main value of analyzing and mining the transaction data of cryptocurrencies is twofold: 1) Transaction records in traditional financial scenarios are relatively unexplored in existing studies as {these} transaction records are usually not publicly accessible for the sake of security and interest. Through analysis and mining of cryptocurrency transaction records, we can extensively explore {\color{black}the} trading behaviors, wealth distribution, and generative mechanism of a transaction system, as well as infer reasons for fluctuations in the financial market of cryptocurrencies. This study can also provide a reference for knowledge discovery in other financial systems. 2) Due to the anonymity of blockchain systems and the lack of authority, various types of cybercrimes have arisen {\color{black}in} the blockchain ecosystem {\color{black}during} recent years. Extracting information from the transaction records can help track cryptocurrency transactions and identify illegal behaviors, thereby establishing effective regulation and building a healthier blockchain ecosystem.

Networks are a general language for describing interacting systems in the real world and complex network science has been widely considered as an effective tool to analyze the modeling, dynamics, and robustness of many networked systems. A considerable part of existing work on cryptocurrency transactions is studied from a network perspective by abstracting objects in cryptocurrency systems such as accounts, smart contracts, and entities as nodes, and the relationship{\color{black}s} between them as links. In a particular cryptocurrency system, there may exist several different interactive activities among users, such as money transfer, smart contract creation, and smart contract invocation.
Networks can be constructed to model these interaction activities on the system from different aspects, and then a variety of network analysis approaches can be employed to analyze network features, extract transaction information, as well as {\color{black}to} detect abnormal or illegal behaviors.

Therefore, as an emerging and interdisciplinary research area, increasing research efforts have been devoted to the analysis and mining of cryptocurrency transactions from a network perspective. Studies in this area not only advance theories and applications of graph data mining techniques on financial systems but also benefit the development of financial security and regulation technologies of blockchain-based cryptocurrencies.
In this paper, we aim to provide a comprehensive review and summary of existing literature and state-of-the-art techniques in this area, with a focus on modeling, profiling, and prediction issues of cryptocurrency transaction networks. In particular, since Bitcoin \cite{nakamoto2019bitcoin} and Ethereum \cite{wood2014ethereum} are the two largest and relatively mature blockchain systems, much of existing research focuses on these two systems.

{\color{black}There are many surveys} about the blockchain technologies and applications in recent literature, including blockchain architecture and technological challenges \cite{bonneau2015sok,zheng2018blockchain,zheng2017overview},
consensus mechanisms \cite{mingxiao2017review}, smart contracts~\
security \cite{li2017survey,DBLP:conf/post/AtzeiBC17,lin2017survey},
anonymity and privacy \cite{khalilov2018survey,feng2019survey},
and blockchain applications \cite{Zhao2016Overview,salah2019blockchain,xie2020blockchain}.
These survey articles aim to discuss some key concepts, technologies, as well as application areas of blockchain but do not intend to provide a thorough summary of the techniques and progress of transaction mining in blockchains.

In \cite{chen2018blockchain}, Chen et al. summarized the types of data in blockchain systems and proposed seven key research problems about blockchain data analysis. Khalilov and Levi \cite{khalilov2018survey} provided a comprehensive overview on analyses of anonymity and privacy in Bitcoin, including studies of entity recognition.
Similarly, Li et al. \cite{li2019survey} presented a survey about anomaly detection in blockchain using data mining techniques. 

Different from the aforementioned surveys, in this work, we aim to present a comprehensive review of state-of-the-art literature on transaction mining of blockchain-based cryptocurrencies conducted from a network perspective. {\color{black}We mainly review the articles published in scientific international journals and proceedings of international conferences from January 2009 to May 2020, and we also include some other kinds of work like book chapters, Masters theses, arXiv papers and technical reports in this survey. We focus our survey on cryptocurrency transaction network analysis by using specific terms including ``cryptocurrency'', ``transaction'', ``network'', ``Bitcoin'', ``Ethereum'' and several popular cryptocurrencies as the search keywords in Google Scholar. Papers not related to cryptocurrency transaction network analysis are not included.}

\begin{figure*}[htbp]
	\centerline{\includegraphics[scale=0.4]{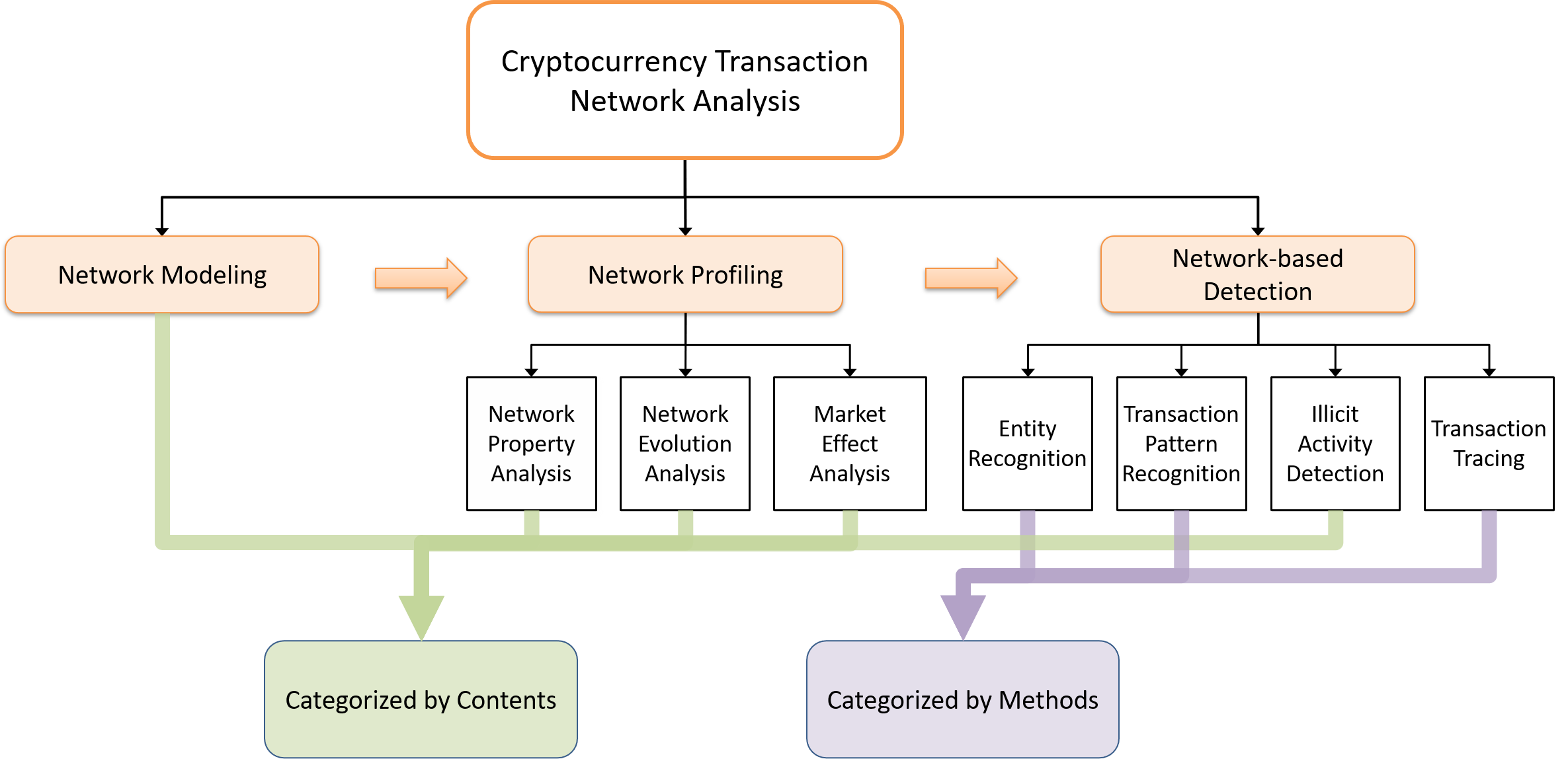}}
	\caption{The outline of cryptocurrency transaction network analysis methods.}
	\label{outline}
\end{figure*}

As shown in Fig. \ref{outline}, we categorize the existing techniques and results on transaction network analysis of cryptocurrencies into three main parts: (\textit{i}) network modeling, (\textit{ii}) network profiling and (\textit{iii}) network-based detection.

In the part of network modeling, we present how to construct a network (graph) to better represent the transaction information for various tasks, and categorize the network modeling methods proposed in existing studies by the semantics of nodes and edges. 

{\color{black}Later in the part of network profiling, we include and introduce existing empirical and analytical studies which focus on extracting descriptive information and providing an overview of cryptocurrency transaction networks. These studies can be further divided into three categories according to their perspectives, namely, network properties, network evolution, and market effects. The studies on network properties mainly focus on measuring or characterizing the transaction network from a static and pure structural perspective, while the studies on network evolution are conducted by incorporating the temporal information and considering the evolution of the networks over time. Furthermore, the analytical studies on market effect discuss the dynamic characteristics of the cryptocurrency market based on the transaction network.}

Lastly, we summarize the techniques and key results of some detective tasks on cryptocurrency transaction networks, such as entity recognition, transaction pattern recognition, illicit activity detection, and transaction tracing. {\color{black}These four tasks are relatively independent but there also exists a progressive relationship between them. First, entity recognition helps cluster pseudonymous addresses sharing the same ownership, which usually is the basis of other downstream detective tasks. Then, detecting illicit activities in blockchain usually has to combine with transaction pattern recognition. Finally, transaction tracing is a more downstream task, such as tracing the money flows involved in the detected illicit activities.}

The main contributions of this work are listed as follows:
\begin{enumerate}[1.]
	\item We provide a detailed and structured overview of the state-of-the-art literature on analyzing and mining cryptocurrency transaction networks, including network construction, network profiling, and network-based detection. 
	\item We summarize the enlightening and important results in existing literature, and we propose several potential research directions for future work on investigating transaction networks of cryptocurrencies.	
	\item To facilitate and foster future research on this topic, we summarize some benchmark data sources of cryptocurrency transaction networks and provide a friendly guidance to new researchers in this area.
\end{enumerate}

The remaining parts of this survey are structured as follows. Section \ref{BG} introduces the background of cryptocurrencies and describes the sources of the datasets. Sections \ref{NM}-\ref{NBD} provide a detailed summary for techniques about network modeling, network profiling, and network-based detection on cryptocurrency transaction networks. Furthermore, we summarize {\color{black}the reviewed studies} and point out several future research directions in Section \ref{FRD}. Finally, we conclude this survey in Section \ref{C}.

\section{Background} \label{BG}
In this section, we {\color{black}introduce the background of} cryptocurrency transaction network analysis. We begin with a brief view of cryptocurrencies, in which we emphatically introduce several mainstream cryptocurrencies. Then we present the transaction process of cryptocurrencies, which is the foundation for further analysis of the transaction data. At the end of this section, we summarize the data sources for both the transaction data and some related label data.

\subsection{Cryptocurrency}
In recent decades, Internet technology has experienced rapid development and gradually gave birth to digital currency. The early form of digital currency can be traced back to the proposal of blind signature technology in {\color{black}the} 1980s and an untraceable payment system based on this technology~\cite{chaum1983blind}. This technology prevented centralized institutions that provide signatures from linking users to their transactions. A series of other digital currency payment technologies like universal electronic cash \cite{okamoto1991universal}, untraceable off-line cash \cite{brands1993untraceable}, fair blind signatures \cite{stadler1995fair}, fair off-line e-cash \cite{frankel1996indirect} later emerged in {\color{black}the} 1990s. However, a common problem existed in these technologies is that\textemdash trusted third parties are needed to detect double spending attacks. 
In the late 1990s, approaches like B-Money \cite{dai1998b}, Bit Gold \cite{szabo2005bit} emerged with an attempt to eliminate the middleman in the trading process. Among them, B-Money first proposed to create coins by solving computational puzzles and decentralized consensus. In a sense, the embryonic form of cryptocurrencies\textemdash virtual currencies secured by cryptography with no link to any central authority appeared in B-Money. However, these approaches ultimately failed to gain enough attention, and the implementation of decentralized consensus has been an unsolvable problem for a long time.

The turning point appeared in 2004 when Hal Finney introduced the idea of ``Reusable Proofs of Work'' (RPoW) \cite{finney2004rpow} which relies on trusted computing as a backend. In 2008, the Bitcoin system, a P2P electronic cash system, was announced by Satoshi Nakamoto. Followed by the deployment of the Bitcoin client in 2009, currency in Bitcoin (bitcoin, abbreviation BTC), the first decentralized cryptocurrency, generated as rewards and transaction fees for Bitcoin miners who create a new block by solving a computationally difficult puzzle. As the basic technology underlying Bitcoin, blockchain has received extensive attention, being widely used in intelligent finance, the Internet of Things (IoT), etc{\color{black}.} \cite{panarello2018blockchain,lu2018blockchain,abbas2019survey}. And in 2017, the price of bitcoins even came up to a peak point of approximately \$20,000 per bitcoin.

After the success of Bitcoin, a number of alternative cryptocurrencies known as ``altcoins'' rapidly emerged. 
As of the second quarter of 2020, there are more than 7,000 kinds of cryptocurrencies with a total market cap of 300 billion dollars\footnote{\url{https://coinmarketcap.com/}}.
Among them, Ethereum \cite{wood2014ethereum} is the largest blockchain system enabling turing-complete smart contracts, and the main currency in Ethereum, which is referred to as Ether (abbreviation ETH), is currently the second-largest cryptocurrency in the world only after {\color{black}B}itcoin.  
One of the earliest altcoins called Namecoin \cite{Vinced2011namecoin} allows users to register with their own domains. Litecoin \cite{lee2011litecoin}, created in 2011, is a kind of cryptocurrency similar to Bitcoin but four times faster than Bitcoin in transaction confirmation. Peercoin \cite{king2012ppcoin}, proposed in 2012, adopts Proof of Stake (PoS) as its consensus algorithm, and PoS is an energy-saving alternative to Proof of Work (PoW) in Bitcoin.
{\color{black}Ripple\footnote{\url{https://ripple.com/}} is a credit network based on distributed open source protocol, it provides a real-time cross-border payment environment that allows transactions between legal tenders and cryptocurrencies with low transaction fees. Because of the successful business model of Ripple, Ripple's XRP token have stood in third place in the cryptocurrency market.}
Other famous cryptocurrencies include Monero \cite{Monero}, Zerocash \cite{sasson2014zerocash}, EOS \cite{xu2018eos}, and Libra \cite{amsden2019libra}, whose detailed information can be found in their white papers.

\subsection{Transaction}
In blockchain systems, a transaction can be regarded as a user operation on the system. When a new transaction is initiated by a user, it will be broadcast to all nodes in the P2P network and added to a new block. 

Transaction models of blockchain systems can be generally categorized into the transaction-centered model and the account-centered model, with Bitcoin and Ethereum being typical examples, respectively \cite{DBLP:journals/corr/abs-1908-04507}. 

In Bitcoin, users are identified by Bitcoin addresses, which are hashes generated from their corresponding public keys. A user can possess multiple addresses to enhance anonymity. The transaction model employed by Bitcoin is a transaction-centered model, where a transaction can have multi-inputs and multi-outputs, being associated with multi-addresses. The inputs are made up of a set of unspent transaction outputs (UTXOs) whose sum of amount is not less than the amount that is to be paid, and the payer can designate a new address to receive the change. Besides, there is no notion for the account balance in Bitcoin. {\color{black}The balance of a Bitcoin user} can be calculated by the sum of the amount value of available UTXOs in its wallet.

The transaction model in Ethereum is {\color{black}an} account-centered model, which contains two kinds of accounts, namely externally owned accounts (EOA) and contract accounts. An EOA is similar to a bank account, which can deposit/draw money, and record some dynamic state information like account balance. Particularly, an EOA can create contract accounts and invoke smart contracts. Each contract account is associated with a piece of executable bytecode and maintains state information like the hash value of the bytecode as well as its account balance. A transaction in Ethereum is a signed data package from an account to another and it contains only one input and one output, which is different from the scenario in Bitcoin. There are three main types of functions that transactions in Ethereum can complete, namely money transfer, contract creation, and contract invocation. {\color{black}According to the type of transaction sender, transactions can be divided into external transactions and internal transactions. A transaction is external only if it is initiated by an EOA, while an internal transaction is triggered by an invocation of a contract and the contract is its transaction sender.} It is worth noting that an external transaction (i.e. a contract function calling) may result in many internal transactions.

\subsection{Dataset}
The entire blockchain data is accessible to the public, thus creating an unprecedented opportunity for transaction network analysis. In this part, we provide a friendly guide on how to collect  blockchain-related data, including transaction records and label information for the cryptocurrency ecosystem. 
\begin{table}
	\renewcommand{\arraystretch}{1.2}
	\caption{\label{Labelwebsite}Services that provide blockchain label information.}	
	\centering	
	\setlength{\tabcolsep}{1.15mm}{	
		\begin{tabular}{|c|p{2.2cm}<{\centering}|p{4cm}<{\centering}|}	
			\hline		
			Category&Name&Introduction\\		
			\hline
			\hline
			\multirow{8}*{Bitcoin}& Walletexplorer \cite{walletexplorer} & A smart blockchain explorer with address grouping, and provides labels for addresses.\\ 
			\cline{2-3}
			& Bitcoin Forum Discussion \cite{bitcointalkTheft} & A discussion of Bitcoin Forum maintains a list of addresses associated with Bitcoin heists, thefts, scams and losses.\\ 
			\hline
			\multirow{6}*{Ethereum} & Etherscan's Label Word Cloud \cite{Etherescanlabelcloud} & Providing labels that tagged by users in Etherscan.\\ 
			\cline{2-3}
			& EtherscamDB \cite{etherscamdb} & Including Ethereum accounts associated with scams, and this website has evolved into CryptoScamDB.\\ 
			\hline
			\multirow{5}*{Others} & CryptoScamDB \cite{cryptoscamdb} &  CryptoScamDB provides an open-source dataset which tracks malicious URLs (e.g., phishing URLs) and their associated addresses for cryptocurrencies.\\ 
			\hline
	\end{tabular}}
\end{table}

By installing a client (such as BitcoinCore\footnote{\url{https://bitcoin.org/en/bitcoin-core/}} for Bitcoin and Geth\footnote{\url{https://geth.ethereum.org/}} for Ethereum) to access the P2P network and then synchronize the block data, we can obtain the raw data from blockchain systems. However, for many of the cryptocurrencies, their raw blockchain data are stored in the binary format and needed to be parsed into human-readable formats for further analysis. Thus we can extract the transaction records from the obtained raw data by building a parser according to the blockchain data structure. Recently, many parsers and modified versions of clients have emerged, e.g., Rusty Blockparser\footnote{\url{https://github.com/gcarq/rusty-blockparser}}, Ethereum ETL\footnote{\url{https://github.com/blockchain-etl/ethereum-etl}}, etc. On the other hand, some clients provide JSON-RPC interfaces for users to obtain the transaction data. Besides, we can crawl the transaction data from block explorer services. Particularly, for some blockchain systems enabling smart contract functionality, the records of internal transactions are not stored in the blockchain data. Yet these internal transaction records can be obtained by utilizing the APIs provided by some block explorer services. We can also obtain the details about internal transactions by replaying all external transactions with a customized client, yet it is a time-consuming process. For Ethereum, the ``trace'' module in the client OpenEthereum\footnote{\url{https://openethereum.github.io/JSONRPC-trace-module}} provides detailed run-time data generated in Ethereum virtual machine, which is a more convenient tool to access the internal transaction records.

{\color{black}The} label information is necessary for some special transaction network mining tasks such as fraud detection and transaction pattern analysis. However, users {\color{black}in} blockchain systems conduct transactions under pseudonyms, making it hard to obtain their identity or label information. Here we summarize some notable websites providing label information in blockchain systems, whose details are shown in Table \ref{Labelwebsite}.

Besides, several well-processed blockchain datasets have also been released for transaction network analysis. Examples include the XBlock-ETH datasets \cite{zheng2020xblock}, the Elliptic dataset \cite{DBLP:journals/corr/abs-1908-02591} for Bitcoin illicit transaction prediction, Bitcoin OTC and Bitcoin Alpha datasets \cite{kumar2016edge}, etc. 

\section{Network Modeling} \label{NM}
\begin{figure*}
	\centering
	\subfigure[Transaction network.]{
		\label{transactiongraph}
		\includegraphics[scale=0.4, trim=-50 0 -100 0]{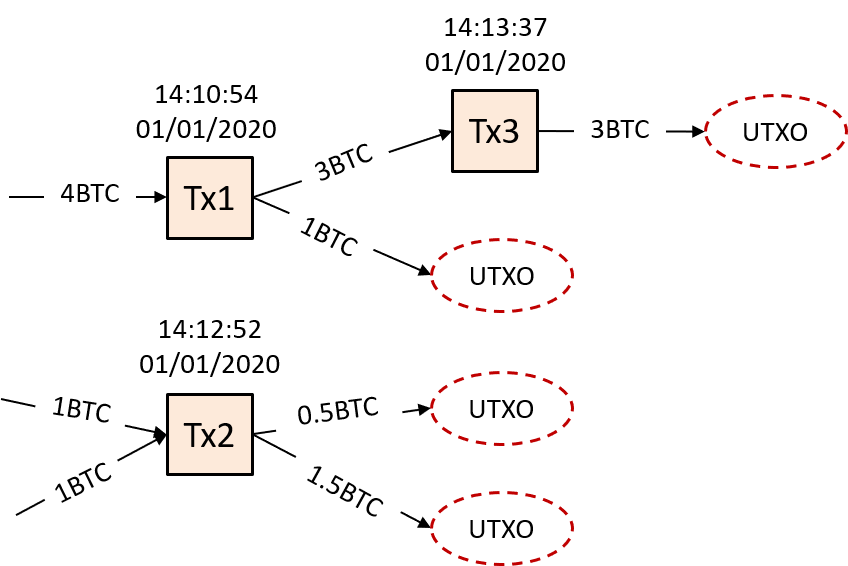}}
	\subfigure[Address network.]{	
		\label{addressgraph}
		\includegraphics[scale=0.4,  trim=-70 -20 -70 0]{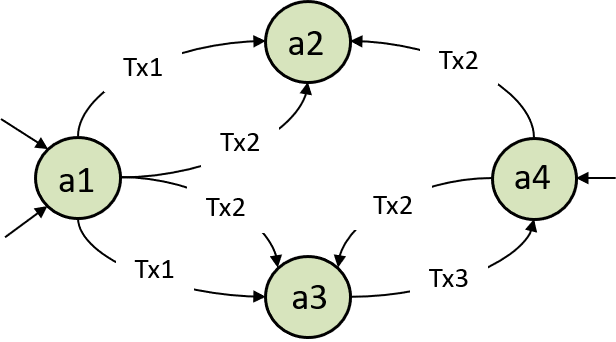}}
	
	\subfigure[User network.]{	
		\label{usergraph}
		\includegraphics[scale=0.4, trim=-90 0 -120 0]{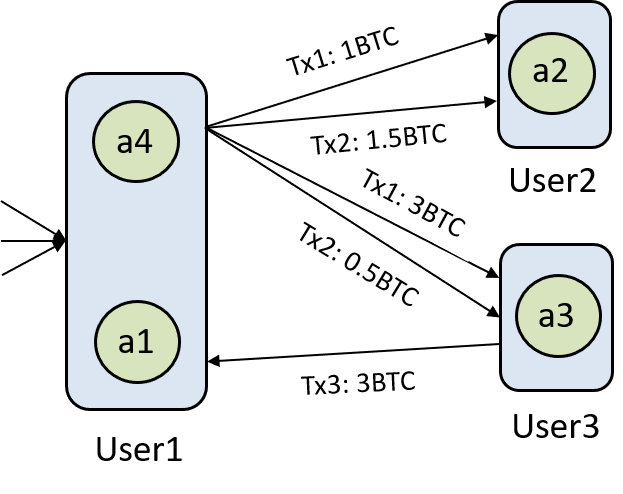}}
	\subfigure[Hypergraph.]{	
		\label{hypergraph}
		\includegraphics[scale=0.4, trim=-40 -20 -20 0]{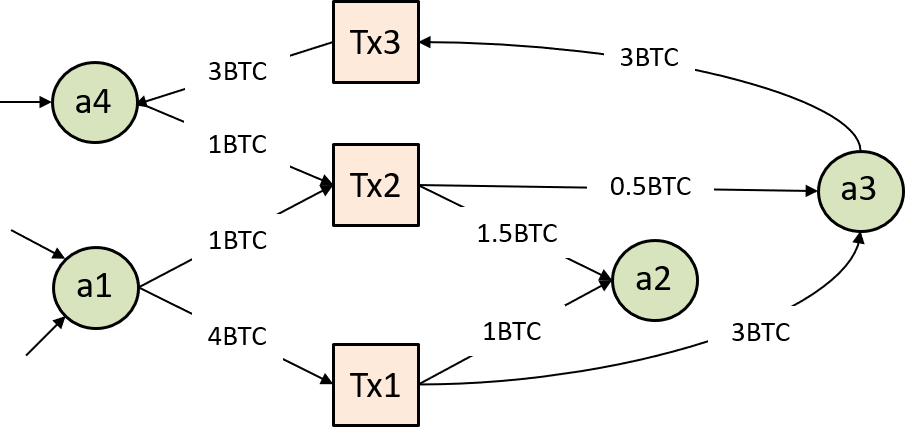}}
	\caption{Examples of network modeling for Bitcoin transaction data. These four kinds of network representations are derived from the same transaction data, where the circular nodes represent Bitcoin addresses, the rounded rectangular nodes represent users, and the rectangular nodes represent transactions.}
	\label{networkmodeling}
\end{figure*}

\begin{figure*}
	\centering
	\subfigure[Money flow graph (MFG).]{
		\label{MFG}
		\includegraphics[scale=0.4]{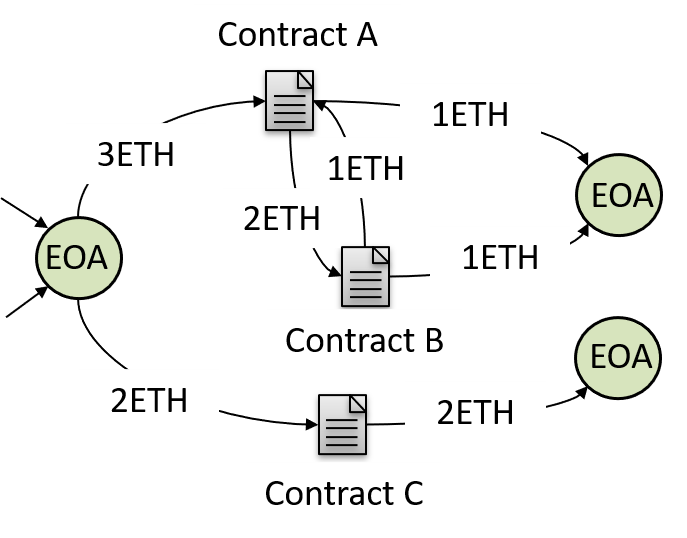}}
	\subfigure[Contract creation graph (CCG).]{	
		\label{CCG}
		\includegraphics[scale=0.4, trim= 0 -50 0 0]{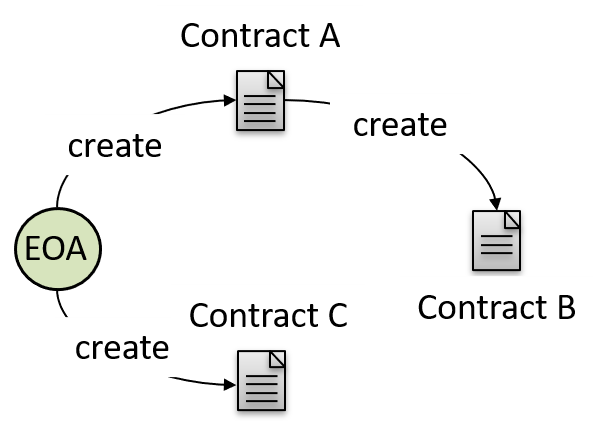}}
	\subfigure[Contract invocation graph (CIG).]{	
		\label{CIG}
		\includegraphics[scale=0.4, trim= 0 -50 0 0]{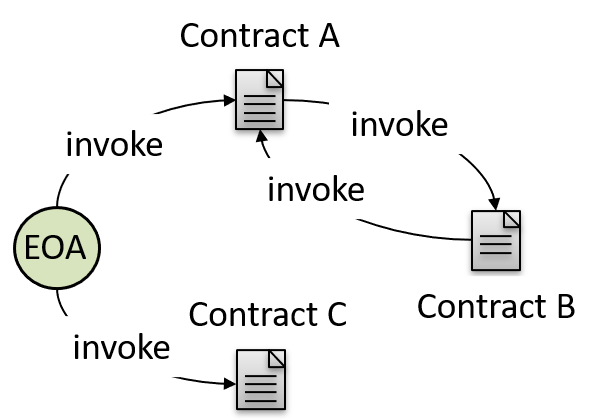}}
	\caption{\color{black}Examples of network modeling for Ethereum transaction data, where the circular nodes represent externally owned accounts (EOA) and the rectangle nodes represent smart contracts.}
	\label{networkmodelingEthereum}
\end{figure*}

After data collection, the first step we need to conduct before {\color{black}cryptocurrency} transaction network analysis is to represent the transaction data into a network-structure data form, namely network modeling.

{\color{black}Different from traditional network modeling process, network modeling for cryptocurrency transaction data has to handle the heterogeneous blockchain data from multiple sources. 
On the one hand, compared with traditional networks with well-defined nodes and links such as citation networks, the definition of nodes and links need to be carefully crafted because of the data heterogeneity.} For example, there are two kinds of accounts in Ethereum, namely EOAs and contract accounts. Besides, transaction relationships in Ethereum contain multiple meanings, such as money transfer, smart contract deployment, and smart contract invocation. Hence it is a challenging work to model {\color{black}the relationships} between objects in blockchain systems. 
On the other hand, {\color{black}blockchain data structure varies from platform to platform}, {\color{black}thus} network modeling for blockchain transaction data is cryptocurrency-oriented.
{\color{black}In general, network modeling methods for cryptocurrencies can be divided into two categories in terms of the transaction model of blockchain systems.}

Fig. \ref{networkmodeling} shows four examples of network modeling for UTXO-based blockchain systems like Bitcoin. For blockchain systems which employ the transaction-centered model, since the inputs of a transaction are UTXOs from previous transactions, the construction of a transaction network is straightforward with transactions as nodes, directed edges including amount and timestamp information representing money flows, as shown in Fig. \ref{transactiongraph}. {\color{black}Reid and Harrigan \cite{Reid2013} are the first to introduce network modeling for cryptocurrency transaction analysis. They proposed the transaction network model (Fig. \ref{transactiongraph}) and the user network model (Fig. \ref{usergraph}) for Bitcoin, representing temporal flows of bitcoins among transaction pairs and user pairs, respectively. Considering a user can own multiple addresses to ensure anonymity, the user network model is based on entity recognition methods (Section \ref{ER}) which can cluster addresses sharing the same ownership, with users as nodes and money flows between user pairs as edges. Since it is difficult to know the true ownership of each address in blockchain systems, the user network model can not achieve completely accurate. These two proposed network modeling methods have been widely adopted in subsequent studies \cite{DBLP:conf/webist/BaumannFL14,pham2016anomaly,meiklejohn2013fistful}.
Another kind of network modeling method is the address network model (Fig. \ref{addressgraph}) introduced in \cite{kondor2014rich}}, where each node represents an address and an edge exists between a pair of addresses if there was at least one transaction between these addresses. Besides, Maesa et al. \cite{DBLP:conf/dsaa/MaesaMR16} modeled the Bitcoin transaction data as a weighted directed hypergraph (Fig. \ref{hypergraph}) with both transactions and addresses as nodes, which can represent the input and output relationship{\color{black}s} between addresses and transactions.

For blockchain systems using the account-centered model like the Ethereum, Chen et al. \cite{chen2018understanding} introduced three kinds of networks for {\color{black}Ethereum transaction data} analysis, namely, money flow graph (MFG), contract creation graph (CCG), and contract invocation graph (CIG). In these three graphs, both EOAs and contract accounts are presented as nodes, but the semantics of edges are different. In MFG, an edge indicates the direction of money flow. While in CCG and CIG, an edge indicates the creation and invocation of a contract, respectively. {\color{black}Fig. \ref{networkmodelingEthereum} shows three examples of network modeling for Ethereum transaction data. As we can see in the figure, the destination of each edge in CCG and CIG is a contract, while the destination of an edge in MFG can be an EOA or a contract. Besides, the edge between two nodes in CCG is unidirectional since a smart contract cannot be created twice. For EOS, which is another popular blockchain platform based on an account-centered model, Zhao et al. \cite{YijingEOS} introduced another two network modeling methods besides MTG and CIG, namely account creation graph (ACG) and account vote graph (AVG). To achieve high throughput and provide an environment for commercial decentralized applications (DApps), EOS adopts the Delegated Proof-of-Stake (DPoS) consensus \cite{dposwhite}. And the voting operations for block producer election in DPoS are stored as a part of transaction actions in EOS. Besides, an account can be created by one existing account in EOS, and the account creation operations are also stored in the form of transaction actions. Hence, ACG and AVG are introduced based on these two EOS's peculiarities and assist to analyze the account creation activities and voting activities in EOS.}



{\color{black}Based on the different semantic definitions of nodes and edges, several network modeling methods are proposed. However,} an important issue is how to make the constructed network better retain the transaction information we need in a specific task. For example, to present the temporal information of transactions, Lin et al.~\cite{lin2020modeling} proposed to model Ethereum transaction records as a temporal weighted multidigraph. Motamed and Bahrak \cite{motamed2019quantitative} introduced the concepts of monthly transaction graph (MTG) and cumulative monthly transactions graph (CMTG), which represent transaction data in a month and the cumulative transaction data, respectively. 
{\color{black}In short, as a fundamental step of network analysis which directly influences the design and the effectiveness of downstream algorithms, modeling the cryptocurrency transaction data with a compatible network model should be a link that can not be ignored.}

\section{Network Profiling} \label{NA}
Up to now, blockchain techniques have been developing for more than ten years. According to CoinMarketCap, there are over 7,000 kinds of active cryptocurrencies, having produced huge transaction data at TB level. Along with the rapid development of blockchain-based cryptocurrencies, a series of concerns have arisen naturally. For example, how do cryptocurrency ecosystems form and develop? Do the cryptocurrency transaction networks have network properties like social networks? Whether the cryptocurrency transactions are in line with the general economic laws? In the past decade, many researchers have devoted to profiling cryptocurrency transaction networks. In what follows, we introduce existing work about network profiling and summarize the related results in terms of network property analysis, network evolution analysis, and market effect analysis.

\subsection{Network Property Analysis} \label{NPA}

\begin{table*}
	\renewcommand{\arraystretch}{1.5}
	\caption{\label{NetworkProperties}Summary of cryptocurrency transaction network properties studied in literature.}	
	\centering	
	\setlength{\tabcolsep}{0.35mm}{	
		\begin{tabular}{|c|p{1.4cm}<{\centering}|p{1.2cm}<{\centering}|p{0.9cm}<{\centering}|p{1.4cm}<{\centering}|p{1.4cm}<{\centering}|p{1.7cm}<{\centering}|p{1.4cm}<{\centering}|p{1.6cm}<{\centering}|p{1.2cm}<{\centering}|}
		\hline
		Property&Number of nodes and edges&Degree distribution &Path length &Clustering coefficient &Centrality &Assortativity coefficient &Connected component &Community &Network motif\\
		\hline
		\hline
		\cite{Reid2013} &\checkmark & \checkmark & \checkmark & & & & \checkmark & & \\
		\hline
		\cite{kondor2014rich} & & \checkmark & & \checkmark & & \checkmark & &  &\\
		\hline
		\cite{ron2013quantitative} & \checkmark &&&&&&\checkmark &&\\
		\hline
		\cite{DBLP:conf/webist/BaumannFL14} & \checkmark &&\checkmark &\checkmark &\checkmark &&&&\\
		\hline
		\cite{lischke2016analyzing} & &\checkmark &\checkmark &\checkmark &\checkmark &&&&\\
		\hline
		\cite{DBLP:conf/dsaa/MaesaMR16} & \checkmark &\checkmark &\checkmark &&\checkmark &&\checkmark &&\\
		\hline
		\cite{DBLP:journals/ijdsa/MaesaMR18} & \checkmark &\checkmark &\checkmark &\checkmark &\checkmark &&\checkmark &&\\
		\hline
		\cite{DBLP:conf/mobisys/JavaroneW18} & &\checkmark &\checkmark &\checkmark &&&&&\\
		\hline
		\cite{alqassem2018anti} & \checkmark &&\checkmark &&&\checkmark &\checkmark &\checkmark &\\
		\hline
		\cite{DBLP:conf/bigdataconf/GaihreLL18} & &&\checkmark &&&&\checkmark &&\\
		\hline
		\cite{chen2018understanding} & \checkmark &\checkmark &&\checkmark &\checkmark &\checkmark &\checkmark &&\\
		\hline
		\cite{DBLP:journals/concurrency/FerrettiD20a} & \checkmark &\checkmark &\checkmark &\checkmark &&&\checkmark &&\\
		\hline
		\cite{guo2019graph} & \checkmark &\checkmark &&&&&\checkmark &&\\
		\hline
		\cite{DBLP:journals/corr/abs-2001-05251} & \checkmark &\checkmark &&\checkmark &&&&&\checkmark \\
		\hline
		\cite{somin2018network} & &\checkmark &&&&&&&\\
		\hline
		\cite{victor2019measuring} & \checkmark &\checkmark &\checkmark &\checkmark &&\checkmark &\checkmark &&\\
		\hline
		\cite{chen2020traveling} & \checkmark &\checkmark &&\checkmark &\checkmark &\checkmark &&&\\
		\hline
		\cite{DBLP:conf/complenet/PopuriG16} & \checkmark &\checkmark &&\checkmark &&\checkmark &&&\\
		\hline
		\cite{liang2018evolutionary} & \checkmark &\checkmark &\checkmark &\checkmark &&\checkmark &\checkmark &&\\
		\hline
		\cite{moreno2018mind} & \checkmark &&&\checkmark &&\checkmark &\checkmark &\checkmark &\checkmark \\
		\hline
		\cite{motamed2019quantitative} & \checkmark &\checkmark &&\checkmark &&\checkmark &&&\\
		\hline
		\cite{YijingEOS}& \checkmark & \checkmark & \checkmark & \checkmark && \checkmark & \checkmark &&\\
		\hline
		\cite{HuangUnderstanding2020}& \checkmark &\checkmark&&\checkmark&\checkmark&\checkmark&\checkmark&&\\
		\hline
	\end{tabular}}
\end{table*}
Complex network theory has been widely demonstrated as a powerful tool in modeling and characterizing various complex systems, including biological systems, transportation systems, social networks, and financial trading systems. To capture particular features of network structure, a {\color{black}variety of} properties {\color{black}and network} measures have been proposed. In the following part, some important network properties and related work studying these properties in cryptocurrency transaction networks are briefly introduced and summarized in Table \ref{NetworkProperties}.

  a) \textit{Number of nodes and edges.} The number of nodes and edges in a network are common measures of network size. Considering these measures, Maesa et al. \cite{DBLP:journals/ijdsa/MaesaMR18} observed that the size of the Bitcoin network increased faster than linear. Alqassem et al. \cite{alqassem2018anti} also reported this phenomenon, and they found that the Bitcoin transaction network is becoming denser and its densification follows a power law rule. Chen et al. \cite{chen2018understanding} presented the number of nodes and edges in MFG, CCG, and CIG of Ethereum, and they found that users transfer money more frequently than calling smart contracts. 
  
  b) \textit{Degree distribution.} In an undirected network, the degree of a node is the number of edges attached to it. Yet in a directed network, a node has two kinds of degrees, namely in-degree and out-degree, which indicate the number of edges pointing into and out from the node, respectively. The degree distribution indicates the probability distribution of node degree in a network, which can be calculated by:
  \begin{equation}
  p(k) = \frac{{\rm the~number~of~nodes~with~degree}~k}{\rm the~number~of~all~nodes}.
  \end{equation}
  One interesting feature in complex network science is that the degree distribution of many realistic systems follows a power-law distribution \cite{DBLP:books/ox/Newman10}, which can be written as:
  \begin{equation}
  p(k) = Ck^{-\alpha},
  \end{equation}
  where $C$ and $\alpha$ are constants and $k$ is the value of the degree. Kondor et al. \cite{kondor2014rich} found that both the in-degree distribution and out-degree distribution of Bitcoin address networks are highly heterogeneous. It has been observed that the degree distribution of many cryptocurrency transaction networks follows a power law. For example, Motamed and Bahrak \cite{motamed2019quantitative} fitted a power-law distribution on the transaction networks for five kinds of cryptocurrencies and found that the power-law parameter $\alpha$ converges to a steady-state after some fluctuations.
  
  c) \textit{Path length.} The path length between two nodes in a network is defined as the minimum number of edges that have to traverse when connecting these two nodes. It is often calculated by the breadth-first search algorithm which can find the shortest path from one node to the other. The diameter of a network is the greatest path length of all the shortest paths in the network. 
  Lischke and Fabian \cite{lischke2016analyzing} showed that the average shortest path of Bitcoin user networks of different countries is in the same range. 
  According to the change of diameter over time, Gaihre et al. \cite{DBLP:conf/bigdataconf/GaihreLL18} deduced the anonymity concerns of users in Bitcoin. 
  Alqassem et al. \cite{alqassem2018anti} analyzed the four possible reasons for the increase in the diameter of Bitcoin transaction networks, namely, anonymity, thieves, the change addresses, and Bitcoin mixing services. 
  Besides, many studies assess the ``small-world'' phenomenon in Bitcoin and Ethereum by calculating the average shortest path \cite{DBLP:conf/mobisys/JavaroneW18,DBLP:journals/concurrency/FerrettiD20a}.
    
  d) \textit{Clustering coefficient.} The clustering coefficient describes the tendency of nodes to gather together in a network. Two measures of the clustering coefficient include the average of the local clustering coefficients and the global clustering coefficient. The local clustering coefficient quantifies the mean probability that two nodes sharing a common neighbor are also themselves neighbors. The global clustering coefficient measures the fraction of paths of length two in the network that are closed \cite{DBLP:books/ox/Newman10} and can be calculated by:
  \begin{equation}
	C_g = \frac{1}{n}\sum_{v}{\frac{t_v}{k_v(k_v-1)/2}},
  \end{equation}  
  where $n$ is the number of nodes, $k_v$ is the degree of node $v$ in the undirected network, and $t_v$ is the number of triangles containing node $v$. 
  Baumann et al. \cite{DBLP:conf/webist/BaumannFL14} observed that Bitcoin user networks have a rather high average clustering coefficient and typical ``small-world'' property.
  Similar results have also been found in the MFG of Ethereum \cite{chen2018understanding}. Yet recent studies revealed that the clustering coefficient of transaction networks of Ripple and Namecoin is relatively low \cite{liang2018evolutionary,moreno2018mind}. 
  
  e) \textit{Centrality.} There are many kinds of centrality measures for networks, which can quantify the importance of a node in a network. The simplest centrality measure is node degree, and other widely considered centrality measures include eigenvector centrality, Katz centrality, PageRank, betweenness centrality, and closeness centrality.
  Lischke and Fabian \cite{lischke2016analyzing} applied the degree centrality to identify major hubs in Bitcoin from September 2012 to April 2013. They found that during this period, exchanges, gambling businesses, and web wallet services were the top major hubs in the Bitcoin ecosystem. Chen et al. \cite{chen2018understanding} listed the top 10 most important nodes in MFG, CCG, and CIG by PageRank. They found that financial applications such as exchanges play an important role in money transfer, contract creation, and contract invocation. For the Ethereum ERC20 ecosystem, top traders selected by PageRank have transactions with each other with a high frequency \cite{chen2020traveling}.
  
  f) \textit{Assortativity coefficient.} The assortativity coefficient measures the tendency of nodes to connect with other nodes in some similar ways. For example, the degree assortativity coefficient implies whether nodes in a network prefer to interact with others having a similar degree. A positive value of the assortativity coefficient with a maximum of 1 implies that the connectivity within the network is assortative mixing, while a negative value with a minimum of -1 implies disassortative mixing, and 0 implies uncorrelation. A series of studies suggested that many blockchain systems like Bitcoin, Ethereum, Litecoin, and Dash are reported to be disassortative in their cryptocurrency transaction network \cite{kondor2014rich,chen2018understanding,motamed2019quantitative}.

  g) \textit{Connected component.} In undirected networks, a connected component is defined as a subgraph where each node can be reached from any others. While in directed networks, the concept of connected component includes weakly connected component and strongly connected component. The definition of weakly connected components is similar to that of connected components in undirected networks. A strongly connected component is the maximal set of nodes where there exists at least one directed path between each pair of nodes. While for calculating the weakly connected components, the edge directions are ignored in the directed networks.
  Statistics and analyses of connected components can help us understand the network structure. Gaihre et al. \cite{DBLP:conf/bigdataconf/GaihreLL18} observed that the number of connected components in the Bitcoin transaction network soared before 2011 but shrank later since many exchanges sprang up around 2010 and 2011, which promoted the circulation of bitcoins. Similar to other networks, most of the Bitcoin addresses are included in the largest connected component (LCC) of the Bitcoin transaction network as reported in \cite{alqassem2018anti}. For Ethereum, Guo et al. \cite{guo2019graph} found that the distribution of the finite component size can be approximated by the power-law model where a heavy-tailed property exists.
  
  h) \textit{Community.} Community is network modules with internally dense connections and externally sparse connections. A network can be partitioned into a given number of communities with community detection algorithms. The partition of different communities presumably reflects the partition of functionality within a network. Alqassem et al. \cite{alqassem2018anti} investigated the properties of time-evolving community structure in Bitcoin. Their study found that the distribution of community sizes can be fitted by the exponentially truncated power law, and the majority of Bitcoin communities have a tree-like structure. Moreno-Sanchez et al. \cite{moreno2018mind} studied how communities are formed in Ripple, they observed that user communities are dynamic and are formed via connecting to gateways in the same geographical region.
   
  i) \textit{Network motif.} Network motifs in complex networks are defined as recurrent subgraph patterns whose occurring number is significantly higher than that in randomized networks. Motifs are an efficient tool to reveal higher-order organizations in networks, and they are well known as the simple building blocks in complex systems. Bai et al. \cite{DBLP:journals/corr/abs-2001-05251} investigated thirteen kinds of 3-node motifs in MFG of Ethereum, and classified these motifs into closed and open triplets, finding that though the number of closed triplets has increased, its proportion shows a decreasing tendency. And the average time for an open triplet to be closed ranges from 37 to 64 days. Paranjape et al. \cite{paranjape2017motifs} observed that the fraction of cyclic triangle motifs is much higher in Bitcoin compared to any other datasets like StackOverflow. Moreno-Sanchez et al. \cite{moreno2018mind} classified wallets into gateways, market makers, and users, and concluded that gateways are the key roles in Ripple from the most frequent motif, which is consistent with the network properties of low clustering coefficient and disassortativity.  
   
  In addition to the well-known network properties discussed above, some researchers have studied cryptocurrency transaction networks from several new perspectives. For instance, unlike other studies focusing on global network properties, Ron and Shamir \cite{ron2013quantitative} investigated the user behaviors such as how they spend bitcoins, how they move bitcoins between their various accounts, and analyzed the largest transactions in Bitcoin. They found that the major of bitcoins remained dormant in addresses at the time of their analysis, and there are many strange-looking structures like binary tree-like structures, long chains in the address network.
  Lischke and Fabian \cite{lischke2016analyzing} conducted an analysis of the Bitcoin user network and economy by cooperating with off-chain data including business tags, IP addresses, and geo-locations. They gave insights into the business distribution as well as the transaction distribution for different countries, and how network properties vary for different subgraphs divided by business types and countries.
  Based on results of prior work~\cite{DBLP:conf/dsaa/MaesaMR16,DBLP:journals/ijdsa/MaesaMR18}, Maesa and Ricci \cite{DBLP:conf/complexnetworks/MaesaMR16} analyzed the outliers of the in-degree distribution in the Bitcoin user network and found some abnormal transaction patterns. 
  Via network analysis, Gaihre et al. \cite{DBLP:conf/bigdataconf/GaihreLL18} answered a question about anonymity and privacy: Whether the Bitcoin users care about anonymity? They found that most users pay weak attention to anonymity, and an important interfering factor is the value of their owning bitcoins.
  Chen et al. \cite{chen2020traveling} conducted a graph analysis to characterize the token creator, token holder, and token transfer activities on the Ethereum ERC20 token ecosystem.
  Liang et al. \cite{liang2018evolutionary} studied some network properties of three kinds of cryptocurrencies, and analyzed their competitive power in terms of these properties.


\subsection{Network Evolution Analysis} \label{NEA}

The cryptocurrency transaction networks are dynamic evolving networks with {\color{black}rapid-}increasing nodes and edges, and it is an interesting problem to investigate how today's giant cryptocurrency transaction networks generate and evolve. 
Existing studies on temporal networks are usually conducted with the form of accumulated networks \cite{kondor2014rich} or snapshots such as monthly {\color{black}networks,} weekly networks{\color{black}, etc.}~\cite{liang2018evolutionary}.

{\color{black}Blockchain systems have passed through different phases during their evolution.}

For Bitcoin, Kondor et al. \cite{kondor2014rich} investigated the Bitcoin transaction data between January 2009 and May 2013, and identified two distinct evolution phases of Bitcoin during this period, namely the initial phase and the trading phase. 
The initial phase lasted until the fall of 2010, during which Bitcoin had low activity and {\color{black}was} mostly used for tests. After that, with the popularity of Bitcoin, bitcoins {\color{black}started being} circulated in the market as a kind of currency and then the Bitcoin system moved on to the trading phase. In \cite{alqassem2018anti}, Alqassem et al. suggested that since late 2014, Bitcoin had entered a new phase with heavy mass media and financial speculation.
Cooperating with {\color{black}address clustering, cluster de-anonymization, and time-evolving analysis on four business categories}, Tasca et al. \cite{tasca2018evolution} identified three evolution phases for Bitcoin over the period between January 2009 and May 2015. The early prototype phase lasted until March 2012 and {\color{black}was} mainly characterized by test transactions among a small number of users, and this period is referred to as ``proof of concept" or ``mining-dominated" phase. Next, the second phase dominated by the early adopters continued through October 2013. Since many gambling services and black markets prevailed during this phase, it {\color{black}was} also called a ``sin" phase or ``gambling/black market-dominated" phase. The third phase called ``maturation" or ``exchange-dominated" phase {\color{black}was} characterized by the maturation and healthy development of the Bitcoin economy.
Similarly, the study \cite{lischke2016analyzing} conducted by Lischke and Fabian displayed the composite of Bitcoin businesses over time, and the result is roughly coincident with \cite{tasca2018evolution}.

For Ethereum, Bia et al. \cite{DBLP:journals/corr/abs-2001-05251} found that the development of Ethereum can be differentiated into three phases, namely ``slow start" phase lasted until March 2017, ``outbreak" phase lasted until June 2018 and ``fever abatement" phase later.

In recent years, a series of studies have been conducted to characterize the evolution of cryptocurrency transaction networks using complex network properties.

To clarify the role of social and anti-social forces in Bitcoin development, Alqassem et al. \cite{alqassem2018anti} quantified the system evolution in terms of several key properties. In the evolution of address categories, they observed that there is a growth trend for intermediate addresses after the beginning of the trading phase. In the evolution of the LCC in Bitcoin, they found that the way how most nodes join in the LCC is similar to other social networks.
In the evolution of community structure, they found that the transaction network tends to have a modular structure, and most of the communities are tree-like structures. Kondor et al. \cite{kondor2014rich} found that the initial phase of Bitcoin can be characterized by large movements in some network properties including heterogeneous in-degree distribution and homogeneous outdegree distribution, while the trading phase can be described by stable network properties, disassortative degree correlations, in-degree distribution, and out-degree distribution. Moreover, they studied the preferential attachment \cite{albert2002statistical} and accumulation of wealth in Bitcoin, finding that linear preferential attachment drives the growth of the network, and sublinear preferential attachment drives the accumulation of wealth. Maesa et al. \cite{DBLP:journals/ijdsa/MaesaMR18} studied the evolution of the user network in terms of several properties, and especially confirmed the ``rich-get-richer" property and highlighted the complex nature of the network. Baumann et al. \cite{DBLP:conf/webist/BaumannFL14} found that the degree distribution in Bitcoin converges to a scale-free network over time.

For Ethereum, Ferretti and D'Angelo \cite{DBLP:journals/concurrency/FerrettiD20a} conducted an evolution analysis with block snapshots. They found that more transactions occurred in the 5,000,000th block than usual and there was a spike in the exchange rate of Ether on that day. To investigate the evolution of Ethereum transaction patterns and {\color{black}the relationship between the network dynamics and the exchange rate of Ether}, Bai et al. \cite{DBLP:journals/corr/abs-2001-05251} studied the evolution of Ethereum on three types of temporal networks. They observed a strong correlation between the size of the user-to-user network and the average price of Ether in a time window, analyzed the macroscopic and microscopic burstiness of transactions, and found that the distribution of wealth in Ethereum is uneven since the beginning.

Besides, some researches \cite{DBLP:conf/complenet/PopuriG16,liang2018evolutionary,motamed2019quantitative} studied the evolutionary dynamics of multiple cryptocurrencies and compared their evolution characteristics. {\color{black}Liang et al. \cite{liang2018evolutionary} analyzed the transaction network dynamics of Bitcoin, Ethereum, and Namecoin. They found that the accumulated networks of these cryptocurrencies do not always densify over time. While for the monthly networks, their degree distribution cannot be well fitted by the power-law distribution. Motamed et al. \cite{motamed2019quantitative} compared the transaction network dynamics of five popular cryptocurrencies, and they found that the growth rate of nodes and edges as well as the network density are related to the cryptocurrency price.}

{\color{black}Future work can be devoted to predicting the development trend of entities and cryptocurrencies according to the evolution characteristics of transaction networks. For example, we can develop a risk analysis tool to predict that whether a centralized exchange on blockchain will abscond with all money by analyzing the transaction pattern dynamics of this exchange.}
\subsection{Market Effect Analysis} \label{MEA}
Since the inception of Bitcoin, blockchain-based cryptocurrencies have been attracting an increasing number of investors and are playing an indispensable role in today's financial market. 
In recent years, a market analysis of cryptocurrencies has become a hot research topic, and a wealth of research efforts have been devoted to the characterization and analysis of the cryptocurrency market. In particular, the exchange rate of cryptocurrencies has been attracting special attention because of its high volatility.

Taking Bitcoin as an example, its price rose to approximately \$20,000 per bitcoin from worthlessness within nine years and fluctuates around \$9,000 per bitcoin when we are writing this survey. Recent years have seen an increasing interest in exploring various issues about the cryptocurrency market.

In this part, we will give a brief review of cryptocurrency market analysis, especially the studies cooperating with transaction network analysis.

\begin{figure*}[htbp]
	\centerline{\includegraphics[scale=0.38]{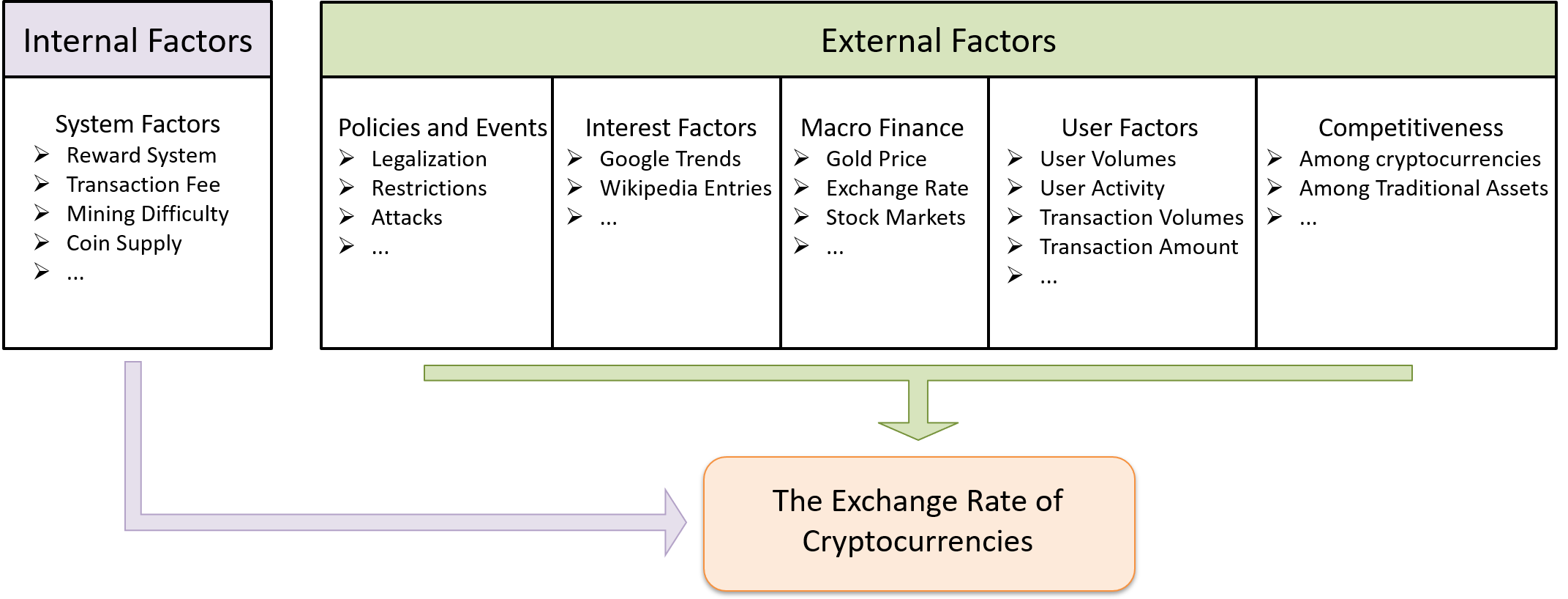}}
	\caption{Factors that influence the exchange rate of cryptocurrencies.}
	\label{factor}
\end{figure*}

Existing studies suggest that the exchange rate or return of cryptocurrencies can be influenced by various factors~\cite{corbet2018exploring,ciaian2016economics,koutmos2018return,sovbetov2018factors}, which can be summarized into internal and external factors (see Fig.\ref{factor}).
Factors directly derived from the system itself are regarded as internal factors, while other factors including policies and events, interest factors, macro-finance, user factors, and competitiveness are regarded as external factors \cite{kristoufek2013bitcoin,krivstoufek2015main,osterrieder2016bitcoin,chen2020dependence,DBLP:journals/sigmetrics/Smuts18}.
Moreover, \cite{koutmos2018return} suggested that cryptocurrencies are becoming more integrated and Bitcoin is the main contributor of their return and volatility spillovers.

Unlike other financial systems, the transaction records of cryptocurrencies are accessible to the public, which also provide an unprecedented opportunity for market effect analysis from the perspective of transaction network analysis. 
Besides, \cite{koutmos2018bitcoin} demonstrated that there is a strong correlation between the transaction activities and exchange rate\textemdash the more users, the more valuable the blockchain ecosystem would become.
Baumann et al. \cite{DBLP:conf/webist/BaumannFL14} noted that some movements in the exchange rate of Bitcoin can be explained by some special events, emphasizing the relationship between user activities and the exchange rate.  
Kondor et al. \cite{kondor2014inferring} captured the change of network structure via principal components analysis on the matrices constructed by the daily network snapshots, and they found the possibility to predict the price of bitcoins with structural changes in the transaction network.  
Yang and Kim \cite{DBLP:conf/ssci/YangK15} examined several properties of the Bitcoin transaction network, and they investigated the dynamic relationship between these properties and market variables like return and volatility.
Bovet et al. \cite{bovet2019evolving} studied the evolution of some network properties in Bitcoin and verified a causal relationship between the exchange rate and the dynamics of network properties.
{\color{black}By using topological data analysis tools, Li et al. \cite{LiDissecting2020} introduced the topology and geometry information of the Ethereum transaction network into blockchain data analysis and showed that the Ethereum transaction network can provide critical insights on forecasting price anomalies and revealing hidden co-movement in pairs of tokens.}

Several studies highlight the effectiveness of higher-order structure in predicting the exchange rate of cryptocurrencies. 
Chen and Ng \cite{DBLP:conf/icdm/ChenN19} proposed a motif-based {\color{black}Long Short-Term Memory} (LSTM) model to predict the price of Storj token in Ethereum by characterizing the dynamics of topological structures in the transaction network.
Akçora et al. \cite{DBLP:conf/pakdd/AkcoraDGK18} proposed the notion of chainlet motifs to characterize topological structures of Bitcoin and {\color{black}their} impacts on the exchange rate. In \cite{akcora2018bitcoin}, chainlet motifs are employed to conduct price prediction and risk modeling of Bitcoin.
However, the original design of chainlet motifs neglected some critical feature information in the transaction networks. {\color{black}Therefore,} Abay et al. \cite{DBLP:conf/icdm/AbayAGKITT19} proposed a feature-preserving representation learning {\color{black}method} on Blockchain transaction networks, {\color{black}and found that transaction network features have a high utility value in predicting Bitcoin price dynamics.}

In addition, Bouoiyour and Selmi \cite{bouoiyour2015does} observed the extremely speculative behavior of Bitcoin, which indicates that investing in Bitcoin is relatively dangerous.
By analyzing the leaked data of Mt.Gox, one famous Bitcoin exchange, Gandal et al. \cite{gandal2018price} pointed out that there existed suspicion of market manipulation in Mt. Gox.
To investigate the existence of market manipulation phenomenons, Chen et al. \cite{DBLP:conf/infocom/ChenWZCZ19} conducted a network analysis on the leaked transaction records of Mt. Gox with singular value decomposition, revealed that many base networks were correlated with the price fluctuation, and observed suspicious market manipulation patterns by further analyzing the behaviors of abnormal users.

{\color{black}The publicly accessible transaction data have provided insights on market effect analysis from a novel point of view. Existing studies in this area are mainly focused on forecasting cryptocurrency exchange rates, revealing hidden co-movement of cryptocurrencies, and modeling cryptocurrency market risk. In fact, there are a lot of research spaces in this area. In particular, nowadays the Decentralized Finance (DeFi) \cite{werner2021sok} ecosystem is mushrooming and has seriously affected the shape of the original cryptocurrency market. However, there is little discussion about the impact of DeFi on the cryptocurrency market. Besides, the publicly accessible transaction network data of decentralized exchanges have produced many interesting and valuable research directions such as washing trading detection \cite{victor2021detecting} and arbitrage analysis \cite{qin2020attacking}.}


\section{Network-based Detection} \label{NBD}
Due to the pseudonymous nature of blockchain technology, the identity information of the users participating in cryptocurrency transactions is usually unknown. Therefore, many blockchain-based cryptocurrency platforms have become a hotbed of various cyber crimes and illegal financial activities.
However, thanks to the openness of blockchain, we can conduct detective tasks on transaction data to identify user identities or abnormal transaction behaviors. Here we categorize network-based detective tasks considered in current {\color{black}literature} into entity recognition, transaction pattern recognition, illicit activity detection, and transaction tracing. 

\subsection{Entity Recognition} \label{ER}
For most blockchain-based cryptocurrencies, the creation of addresses or accounts is very easy and almost cost-free, and thus one particular user may own multiple addresses or accounts to enhance anonymity. 
Ron and Shamir proposed to use the neutral word ``entity'' to describe the common owner of multiple addresses (accounts) \cite{ron2013quantitative}.
Many cybercrimes on cryptocurrency platforms are found to be related to entities with a large number of accounts.
For example, money laundering in blockchain systems is always conducted with multiple addresses or accounts. Therefore, entity recognition, which refers to de-anonymize the entities that own at least one address (account), is regarded as the basis of detective tasks on cryptocurrency transaction networks from the entity perspective. Existing methods of entity recognition can be roughly categorized into three types, namely, transaction property-based,  behavior-based, and off-chain information-based methods.

a) \textit{Transaction property-based methods.} This kind of methods utilizes transaction properties to recognize the accounts belonging to the same entities. For the Bitcoin system which is based on the UTXO model, Reid and Harrigan~\cite{Reid2013} proposed an intuitive identity method assuming that the input addresses of a particular transaction are possessed by the same entity, and this method is referred to as the multi-input heuristic method. 
To spend bitcoins on an address in the Bitcoin system, it is necessary to provide the private key of the address, and usually, Bitcoin users do not share their private keys, so it can be considered that the addresses of the input end of a transaction are under the control of the same entity. The fact that inputs of multi-input transactions may belong to the same entity was also noted by Nakamoto \cite{nakamoto2019bitcoin}.

{\color{black}Harrigan and Fretter \cite{harrigan2016unreasonable} further investigated the reasons behind the effectiveness of this heuristic method. They found that some factors such as address reuse, avoidable merging, super clusters with high centrality and the incremental growth of address clusters may cause the multi-input heuristic method to produce false positives.}
However, the multi-input heuristic method may lead to two kinds of errors \cite{ron2013quantitative}, namely, the underestimation errors and the overestimation errors. The underestimation errors are caused by the neglect of the common ownership among some addresses, while the overestimation errors are caused when multiple users take part in the inputs of a transaction.

By combining both the multi-input heuristic and community detection method, Remy et al. \cite{remy2017tracking} proposed a new entity recognition method, which can increase the recall at the cost of precision, and adjust the recall depending on practical applications. 

Another typical method utilizing transaction properties for Bitcoin is the change address heuristic method which was first mentioned by Reid and Harrigan \cite{Reid2013}. During the transaction process of Bitcoin, change addresses will be automatically generated for the payers to receive the change. Therefore, the change address among the outputs of a transaction can be linked to the entity that owns the input addresses.

Androulaki et al. \cite{androulaki2013evaluating} applied the change address heuristic method for entity recognition in their experiments.
Then Meiklejohn et al. \cite{meiklejohn2013fistful} expanded this method to make it not limited to transactions with two outputs based on the assumption that a change address only has one input.
Besides, Ortega \cite{ortega2013bitcoin} supposed that the change address is the output address with more decimals in the transaction, since in most cases the decimals of the real outputs are reduced.
Nick \cite{nick2015data} proposed an optimal change heuristic with the assumption that the change output value is more likely to be smaller than any of {\color{black}the} other inputs.

Klusman and Dijkhuizen \cite{klusman2018deanonymisation} pointed out that the multi-input heuristic and change address heuristic methods are not applicable to some blockchain systems like Ethereum which are based on the account-centered model. To address this problem, Victor \cite{victoraddress} proposed three heuristic rules based on the facts of deposit account reuse, airdrop multi-participation, and self-authorization for Ethereum. The deposit account reuse heuristic rule is designed in the scenario of exchange, and exchanges typically generate the deposit accounts for users to deposit money. Once users transfer money to their deposit accounts, the money will later be transferred to a hot wallet belonging to the exchange automatically. The deposit account heuristic was developed based on the reuse of deposit accounts by clustering the accounts using the same deposit account into an entity. As for the airdrop multi-participation heuristic, it is designed in the scenario of airdrop, which is a popular advertising method implemented by distributing tokens for fund raising in an initial coin offering (ICO). Since some users may sign up multiple accounts to participate in airdrops and aggregate the tokens to one account, the airdrop heuristic utilizes this fact to conduct entity recognition. The approval heuristic is based on assumption that the token spenders and the owners are the same entity in authorization activity via the call of approve function. Among these three methods, Victor concluded that the deposit account heuristic is the most effective method via experiments.

{\color{black}Besides, Moreno-Sanchez et al. \cite{MorenoSanchez2016ListeningTW} found it possible to deanonymize the identities of wallets in Ripple, and they proposed two heuristics. By observing the sender wallet and the receiver wallet of a deposit (withdrawal) operation in Ripple, the first heuristic can link the Ripple wallet of a user to another wallet (such as Bitcoin wallet) belonging to the same user. Users in Ripple always use hot wallets to perform their daily transactions, while the cold wallets, with sighing keys securely kept in offline devices, are always used as reserves. Since the cold wallets are publicly linked to their users and only send transactions to their related hot wallets, the second heuristic is proposed by linking Ripple wallets with their cold wallets.} 

b) \textit{Behavior-based Methods.} From a cryptocurrency transaction network, transaction features that reveal behavior preferences of users can be extracted, such as the most frequent transaction patterns, average transaction amount, and transaction frequency. Researchers proposed a series of methods to cluster addresses according to the similarity of transaction behaviors. 
Reid and Harrigan \cite{Reid2013} proposed to identify entities by considering the similar transaction times over an extended timeframe. 
Androulaki et al. \cite{androulaki2013evaluating} considered several features of transaction behavior including the transaction time, the index of senders and receivers, and the amount value of transactions, and found that clustering methods based on these features can reveal the identities of almost 40\% of the users. 
Huang et al. \cite{huang2017behavior} utilized the sequences of transaction amount change to develop a behavior pattern clustering algorithm for entity recognition.
Monaco \cite{monaco2015identifying} proposed several features to capture the transaction behavior dynamics of Bitcoin users and found that the behavioral patterns observed over a {\color{black}period} of time can reveal the identity of the users.
Zhang et al. \cite{zhangbitscope} introduced a multi-resolution clustering system for Bitcoin address de-anonymization. 

Some researchers treated the entity recognition problem as a classification problem with transaction behavior features as the input. 
Jourdan et al. \cite{jourdan2018characterizing} explored five types of features including address features, entity features, temporal features, centrality features, and motif features, and studied the efficacy of these features in classifying Bitcoin addresses. 
Harlev et al. \cite{harlev2018breaking} considered transaction features in a supervised machine learning framework to de-anonymize Bitcoin addresses.
Moreover, Shao et al. \cite{shao2018identifying} embedded the transaction history of each Bitcoin address into a lower dimensional feature vector with {\color{black}a} deep learning method, and then applied it to achieve entity recognition. 

c) \textit{Off-chain information-based methods.} Off-chain data refer to the blockchain-related data that are not stored in the blockchain, which can also be used to assist the de-anonymization process. Typical off-chain data include but are not limited to the IP addresses of nodes, the leaked data from exchanges, the business labels. Many services like exchanges, mixing services, and denotation organizations can obtain off-chain information like the IP addresses, e-mail addresses of pseudonymous blockchain users. Some users posted the addresses (accounts) of theft cases and scams they had experienced to forums, providing us an opportunity to get access to the information by crawling the web. If the address information of a user is exposed, we can easily reveal the transaction behaviors and other addresses belonging to this user.

Reid and Harrigan \cite{Reid2013} first applied this method in entity recognition and utilized the off-chain information to identify some entities involved in a theft case. 
According to the off-chain information from Bitcoin forums, Fleder et al. \cite{fleder2015bitcoin} linked the Bitcoin addresses to real people and found some forum users had transactions with Silk Road and gambling services.
Jawaheri et al. conducted a similar investigation to deanonymize hidden service users by exacting information from posts on social media \cite{al2020deanonymizing}.
M\"{o}ser et al. \cite{moser2013inquiry} traded with three kinds of mixing services and recorded the related Bitcoin addresses. Then the address information was used in a mixing service recognition model \cite{DBLP:conf/ciarp/Prado-RomeroDA17}.
Ermilov et al. \cite{ermilov2017automatic} proposed a clustering method for entity recognition, which cooperates with both on-chain information and off-chain information collected from 97 sources.  
In \cite{DBLP:conf/webist/BaumannFL14}, Baumann et al. investigated the IP addresses of Bitcoin users.
Neudecker and Hartenstein \cite{neudecker2017could} suggested that network information can only recognize a small number of entities in Bitcoin,
and Biryukov et al. \cite{biryukov2014deanonymisation} proposed a method to de-anonymize the Bitcoin clients by linking IP addresses to Bitcoin wallet addresses. 

{\color{black}In summary, entity recognition methods are essentially clustering methods that can gather addresses belonging to the same user. Among the above methods, methods based on the transaction properties of a specific system can satisfy most of the transaction rules in the system, and therefore they are effective in most cases. The behavior-based methods can gather the addresses behaving similarly, however, they are relatively coarse-grained for the tasks of entity recognition. By combining with the off-chain information, address clusters can be associated with their corresponding real identities and achieve de-anonymization.}

\subsection{Transaction Pattern Recognition} \label{TPR}
In cryptocurrency ecosystems, transaction behaviors vary from user to user. For example, an exchange entity tends to interact more frequently with other users than a normal entity. Transaction pattern recognition aims to reveal some special transaction network structures and further analyze the users' behaviors. Here we summarize the related studies into three categories which are given as follows:

a) \textit{Visualization methods.} Visualization is a powerful tool for network analysis, and the transaction patterns can be directly observed from the visualization result. By visualizing the Bitcoin transaction networks in some specific blocks, McGinn et al. \cite{mcginn2016visualizing} discovered some unexpected transaction patterns such as the money laundering pattern and the denial of service attacks. 
When investigating the market manipulation phenomenon in Bitcoin, Chen et al. \cite{DBLP:conf/infocom/ChenWZCZ19} visualized the daily subgraphs of abnormal addresses and found some abnormal transaction patterns which are associated with market manipulation, such as self-loop, bi-direction, triangle, and so on.
Ferrin \cite{ferrin2015preliminary} categorized and visualized some observed transaction patterns in Bitcoin, and found that some of these transaction patterns are typical patterns occurring in businesses related to exchanges and mixing services. 
Meiklejohn et al. \cite{meiklejohn2013fistful} discovered a special transaction pattern called ``peeling chain" from many criminal activities in Bitcoin. This kind of transaction pattern starts from an address with a large amount of money, and sequentially creates a new transaction that transfers a small amount of money to one address (as the input of the next transaction) and uses a one-time change address to receive the remainder. This process will be repeated until the money is pared down.
McGinn et al. \cite{mcginn2018towards} visualized the source and destination blocks of Bitcoin transaction flows as an adjacency matrix. They found that this kind of representation can easily reveal some repeated transaction behaviors (also called ``DNA sequences"), which can help associate a transaction with some other transactions having similar behaviors.

b) \textit{Tracking analysis.}
{\color{black}Tracking and observing the transactions of specific addresses can also provide insights into the preferred transaction patterns of these addresses.}
Maesa et al. \cite{maesa2017detecting} analyzed the outliers in the in-degree distribution of the Bitcoin user network and noticed an unusual kind of transaction pattern called the pseudo-spam transaction. After further analysis, they suspected that the pseudo-spam transactions may be part of a user pseudonymity attack or a spam attack, or may possibly be used for advertising.
M\"{o}ser et al. \cite{moser2013inquiry} gained insights into the operation modes of three mixing services by making transactions with them and tracing the following transactions.
Tasca et al.\cite{tasca2018evolution} studied the transaction patterns of exchanges, mining pools, gambling, and black markets in Bitcoin by analyzing their inflows and outflows.
Ron and Shamir \cite{ron2013quantitative} traced the flow of large transactions whose transaction amount is larger than 50,000 BTC in Bitcoin and analyzed the hidden behavior under the detected long chain and fork-merge patterns.

c) \textit{Motif analysis.}
A series of studies proposed to mine the transaction patterns of cryptocurrencies via analyzing network motifs in the transaction network. 
Ranshous et al. \cite{DBLP:conf/fc/RanshousJKNSWW17} represented the Bitcoin transaction network as a directed hypergraph and introduced motifs in directed hypergraphs to reveal the transaction patterns of exchanges.  
Wu et al. \cite{DBLP:journals/corr/abs-2001-05233} proposed the concept of attributed temporal heterogeneous motifs in a directed and temporal Bitcoin transaction network and applied it to detect the addresses belonging to Bitcoin mixing services. 
Using network motifs as features, Zola et al. \cite{zola2019bitcoin} developed a method for entity classification in Bitcoin. They compared the similarities of entity transaction patterns over time and investigated whether some transaction patterns were repeated in different batches of Bitcoin transaction data.
Jourdan et al. \cite{jourdan2018characterizing} applied network motifs to reveal the information of transaction patterns in entity classification, and they found that the transaction patterns can be viewed as the fingerprint of entities. 

{\color{black}In short, transaction pattern recognition is mainly to summarize some common transaction patterns in blockchain systems, as well as to make some discoveries on the transaction patterns of special addresses, which can pave the way for some detective tasks such as illicit activity detection. Though existing work on transaction pattern recognition is mainly focused on Bitcoin-like blockchain platforms, all these mentioned methods are also suitable for other blockchain platforms. In the future, more efforts are needed to conduct transaction pattern recognition on different blockchain systems.}

\subsection{Illicit Activity Detection} \label{IAD}
One special characteristic of blockchain systems is that they are built with pseudonyms, and the users can transact without exposing their real identities. The pseudonymous nature brings cryptocurrencies not only a large number of investors but also a bad reputation in attracting many illicit activities like scams, black markets, money laundering, and so on.
Different from traditional financial scenarios, it is unlikely to enforce Know-Your-Customer (KYC) processes to verify the identities {\color{black}and ascertain the potential risks of users before conducting a cryptocurrency transaction}.
Fortunately, the public and irreversible transaction records provide us an opportunity to detect irregular transaction patterns in blockchain systems. 
 
As openness and irreversibility are also major features of blockchain technology, extracting information from the public and irreversible transaction records is an intuitive and effective way to detect illicit activities.
Most of the existing studies explored the issue of illicit activity detection via anomaly detection by incorporating hand-crafted features or automatically extracted features. In the following part, we will focus on the research work on the detection of financial scams and money laundering, and then give a brief review on the detection of other illicit activities on cryptocurrency transaction networks.

a) \textit{Scams.} Scams are traps designed with fraudulent intent. Existing financial scams in blockchain systems have brought a huge threat to the transaction security and healthy development of the cryptocurrency ecosystem.
Vasek and Moore \cite{vasek2015there} summarized a list of various scams in Bitcoin and conducted an empirical analysis on these scams. They classified these scams into four categories, namely, Ponzi schemes, mining scams, scam wallets, and fraudulent exchanges, and found that 13,000 potential victims had lost approximately \$11 million in 192 scams.
Moreover, some other kinds of scams such as scam Initial Coin Offerings (ICO) \cite{DBLP:journals/corr/abs-1803-03670}, smart contract honeypots \cite{DBLP:conf/uss/TorresSS19} have been found in blockchain systems. {\color{black}The means of these scams are emerging in an endless stream and develop rapidly.}
Therefore, a wealth of research efforts have been devoted to detecting scam activities with blockchain-based cryptocurrencies, and here we mainly review the network-based detection methods.

{\color{black}Based on the hand-crafted features extracted from the transaction network}, Pham and Lee \cite{pham2016anomaly} applied the trimmed $k$-means algorithm \cite{cuesta1997trimmed} to detect the fraudulent activities in Bitcoin.
{\color{black}Similarly, Chen et al.\cite{DBLP:conf/www/ChenZCNZZ18,DBLP:journals/access/ChenZNZZ19} proposed a method for Ethereum Ponzi scheme detection by utilizing the hand-crafted features. What is distinctive is that they not only the account features extracted from the transaction network, but also utilized the code features extracted from the opcodes, making it possible to detect Ponzi scheme contracts at the time of being created.}

{\color{black}Some studies modeled the scam account detection problem as a node classification problem in a transaction network, and use network embedding methods to automatically extract features from the network. Chen et al. \cite{ChenLiangPhishing2020} use Graph Convolutional Network (GCN) to detect phishing scam account in Ethereum. They found that GCN-based methods can achieve a higher detection rate than traditional methods based on hand-crafted features.} Tam et al. \cite{DBLP:journals/corr/abs-1906-05546} proposed a GCN-based method called EdgeProp to learn the embeddings of nodes and edges in large-scale transaction networks. Unlike traditional GCN-based methods, EdgeProp incorporates edge attributes and performs effectively in identifying illicit accounts and capturing the transaction pattern relationships with an Ethereum dataset. 
Besides, Wu et al. \cite{10.3389/fphy.2020.00204,DBLP:journals/corr/abs-1911-09259} proposed two kinds of random walk-based embedding methods that consider some specific features of transaction networks such as transaction amount, timestamp, and multi-edge. With the learned embeddings as node features, the effectiveness of these two methods has been demonstrated in downstream phishing detection tasks.

b) \textit{Money laundering.}
According to \cite{bryans2014bitcoin}, the process of money laundering is usually accomplished with three steps: (1) Firstly, the ``dirty money'' is injected into the financial system. (2) Secondly, the ``dirty money'' is mixed with ``clean money'' and dissociated from its illegal source. (3) Finally, the ``dirty money'' is integrated and repossessed by the criminals in a seemingly legal state.  
Due to the pseudonymous nature and many available privacy-enhancing techniques in blockchain systems, cryptocurrencies have become an optional currency for the process of money laundering to conceal illicit currency flows. According to a report~\cite{fanusie2018bitcoin} about Bitcoin money laundering from Elliptic, which is a famous cryptocurrency intelligence company, exchanges, mixing services and gambling sites are three notable Bitcoin laundering destinations. Meanwhile, the percentage of all incoming transaction volume coming from illicit entities in mixing services is much higher.
Hence in recent years, studies about money laundering detection on cryptocurrency transaction networks mainly focused on detecting mixing services and investigating suspicious transaction patterns of exchanges. 

Mixing services are designed to enhance the privacy of transactions and make the relationships between senders and recipients more untraceable {\color{black}\cite{Ruffing2014Coin,moreno2017pathshuffle}}. To study how mixing services work, M\"{o}ser et al. \cite{moser2013inquiry} investigated the operation models of three mixing services with reverse-engineering methods and tried to trace the transactions back to the origins. 
Prado-Romero et al. \cite{DBLP:conf/ciarp/Prado-RomeroDA17} first proposed the problem of detecting mixing services and tackled it as a community outlier detection problem. They emphasized the importance of mixing service detection as that once the mixing services are detected, we can further analyze whether the addresses or accounts that interacted with these services have taken part in illicit activities. 
However, this work lacks generalization for different kinds of mixing services. To deal with this problem, Wu et al. \cite{DBLP:journals/corr/abs-2001-05233} proposed a feature-based detection method with hybrid network motifs, which can better characterize the transaction patterns of different mixing services.

\begin{table*}
	\renewcommand{\arraystretch}{1.5}
	\caption{\label{TaintAnalysis} Description and evaluation of five taint analysis methods.}	
	\centering	
	\setlength{\tabcolsep}{0.9mm}{	
		\begin{tabular}{|c|p{0.72\columnwidth}|p{0.72\columnwidth}|c|}	
			\hline		
			Taint analysis method& \multicolumn{1}{c|}{Description} & \multicolumn{1}{c|}{Evaluation} & Example\\		
			\hline
			\hline
			\multirow{3}*{Poison} &The outputs of a transaction are considered to be tainted as long as the transaction has at least one dirty input.&The clean money involved in the transaction will be misclassified as the dirty one, and the amount of tainted money will increase exponentially.&\multirow{3}*{Fig.\ref{poison}}\\
			\hline
			\multirow{3}*{Haircut} &By taking the amount value of the dirty inputs into consideration, each output in a transaction contains the proportion of the dirty and clean inputs.&It does not affect the amount of clean money, but results in the exponential growth of tainted transactions and the mixing between clean and dirty money.&\multirow{3}*{Fig.\ref{haircut}}\\
			\hline
			\multirow{3}*{FIFO} &The order of the outputs in a transaction is according to the chronological order of the inputs.&It is a more precise way compared to the Poison and Haircut methods, but it cannot handle the accuracy problem since the order may be inaccurate in some cases.&\multirow{3}*{Fig.\ref{FIFO}}\\
			\hline
			LIFO &Similar to FIFO but has a reversed order.&Similar to the evaluation of FIFO.&Fig.\ref{LIFO}\\
			\hline
			\multirow{3}*{TIHO} &The dirty inputs distribute to the outputs with higher value, while the clean inputs are linked to the outputs with a small value which can be seen as the change outputs.& It aims to capture complex transactions that others cannot accomplish, but it will be invalid if the tainted outputs are smaller than the change. Its accuracy is similar to other methods while having a difference in detected addresses.&\multirow{3}*{Fig.\ref{TIHO}}\\
			\hline
	\end{tabular}}
\end{table*}

{\color{black}For money laundering pattern capturing,} Hu et al. \cite{DBLP:journals/corr/abs-1912-12060} characterized the transaction patterns of Bitcoin money laundering with feature analysis, and they developed classifiers to detect money laundering transactions by employing some network embedding methods like deepwalk \cite{DBLP:conf/kdd/PerozziAS14} and node2vec \cite{DBLP:conf/kdd/GroverL16}.
Battista et al. \cite{di2015bitconeview} proposed the notion of purity which allows the understanding of when and how bitcoins are mixing and built up a system named BitConeView for visual analysis of transaction flows.
Ranshous et al. \cite{DBLP:conf/fc/RanshousJKNSWW17} pointed out that the exchanges provide connections between pseudonyms and real-world identities, and thus studying the transaction patterns of exchanges is an important step for anti-money laundering. With the designed network motifs, they identified addresses being owned by exchanges and characterized the inter-exchange activity.  
Besides, McGinn et al. \cite{mcginn2016visualizing} presented a visualization system to discover and illustrate specific transaction patterns like money laundering operations.

c) \textit{Others.}
Besides financial scams and money laundering, researchers have proposed network-based methods to detect addresses or accounts involved in other illicit activities or misbehaviors.

Since the widespread use of {\color{black}cryptocurrencies} in ransomware payments, Akcora et al. \cite{Akcora2019} proposed a topological information-based ransomware detection framework to find out the addresses associated with both the known ransomware families and new appeared ransomware families.
Conti et al. \cite{DBLP:journals/compsec/ContiGR18} studied recent ransomware and their economic impact from a Bitcoin payment perspective. Incorporating the information of co-input transactions and change addresses, they proposed two clustering heuristics to identify the addresses associated with ransomware.

For the black market investigation, Foley et al. \cite{foley2019sex} quantified and characterized the illicit trading activities in Bitcoin, and they proposed two methods to identify these activities. The first one is that applying a community detection method to identify two distinct communities, while the second one detects illegal users by exploiting characteristics. 
Moreover, the authors suggested that approximately one-half of Bitcoin transactions (46\%) are associated with illegal activity. 

{\color{black}To detect bot accounts in EOS which are operated by machines, Huang et al. \cite{HuangUnderstanding2020} proposed two bot account detection methods based on the transaction network data from two levels, namely the community level and the pre-account level. Based on these two methods, they identify 381,837 bot accounts in EOSIO in total. These bot accounts have a high possibility to be involved in illicit activities such as manipulating the DApp ranking list.}

Weber et al. \cite{DBLP:journals/corr/abs-1908-02591} contributed an Elliptic dataset which is a Bitcoin transaction network with over 200 thousand transactions as nodes, 234 thousand transaction flows as edges and 166 node features. Transactions in this dataset are labeled into the licit category (e.g. exchanges, wallet providers, miners, and licit services), illicit category (e.g., scams, malware, terrorist organizations, and ransomware), and unlabeled category according to real entity information. Based on this dataset, the authors presented the detection results using several methods including the temporal EvolveGCN \cite{pareja2019evolvegcn}.

{\color{black}All in all, in recent years many illicit activity detection methods have been proposed for blockchain. However, most of these techniques can only report the detected illicit activities after these illicit activities have happened, which are lacking in early warning and intelligent interception of illegal transactions.}

\begin{figure*}
	\centering
	\subfigure[Poison method.]{
		\label{poison}
		\includegraphics[scale=0.4]{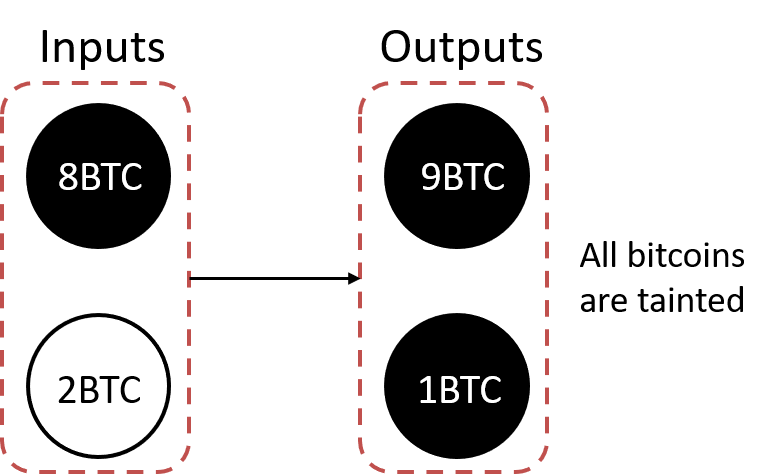}}
	\subfigure[Haircut method.]{	
		\label{haircut}
		\includegraphics[scale=0.4, trim=0 18 0 0]{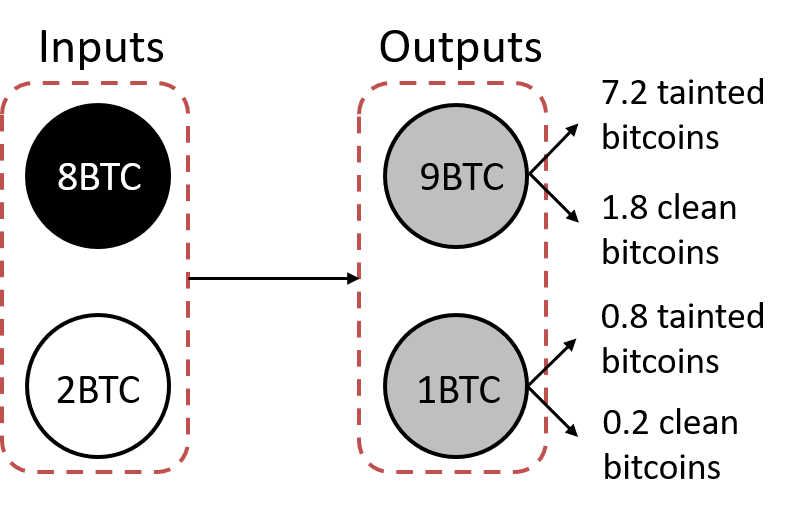}}
	
	\subfigure[FIFO.]{	
		\label{FIFO}
		\includegraphics[scale=0.4]{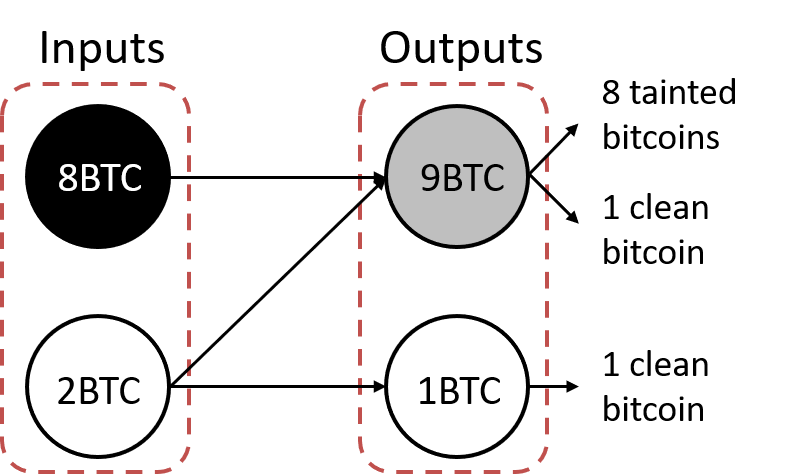}}
	\subfigure[LIFO.]{	
		\label{LIFO}
		\includegraphics[scale=0.4]{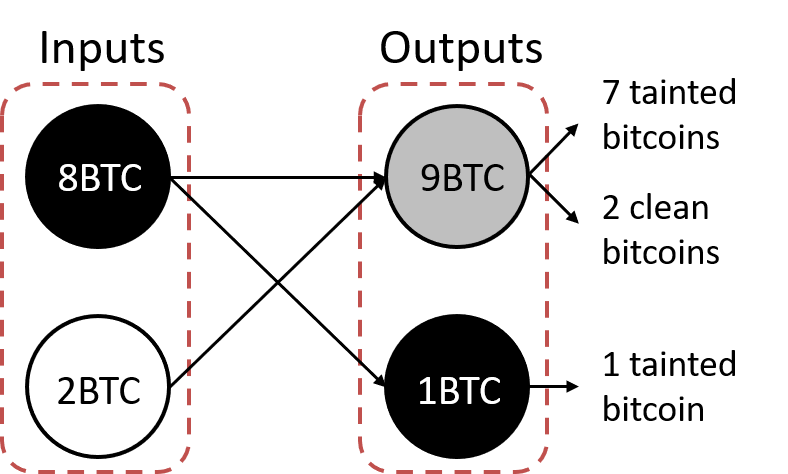}}
	\subfigure[TIHO.]{	
		\label{TIHO}
		\includegraphics[scale=0.4]{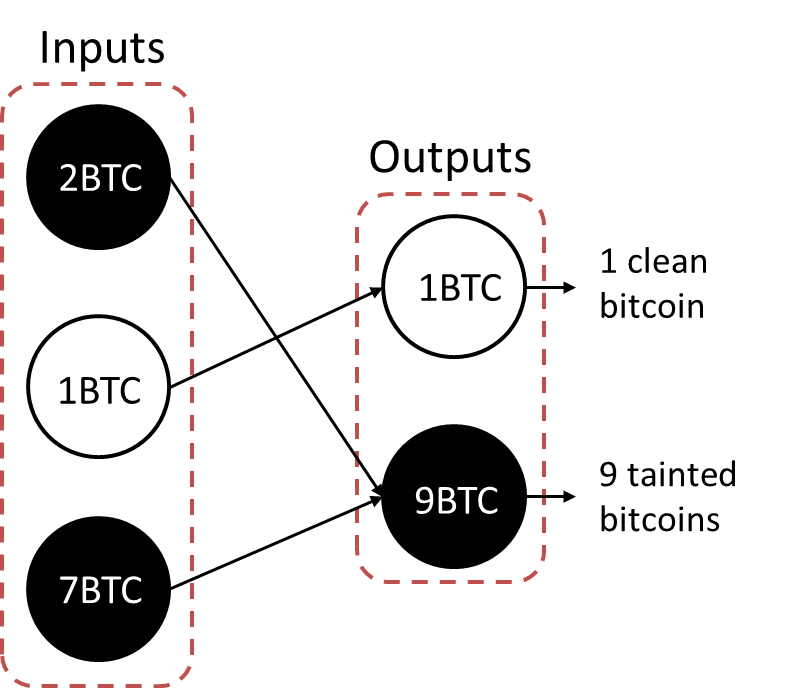}}
	\caption{Examples for different taint analysis methods. Each figure shows the inputs, outputs of a certain transaction, and the money flow directions between them. For the inputs, a white circle represents an input with clean bitcoins, while a black circle represents an input with tainted bitcoins. And for the outputs, the color of circles implies the taint analysis result, where the white circle represents the output bitcoins are clean, the black circle represents the output bitcoins are totally tainted, and the gray circle represents the output bitcoins are a mixture of both tainted and clean bitcoins.}
	\label{taintanalysis}
\end{figure*}
\subsection{Transaction Tracing} \label{TT}
Recently, the rise and development of transaction obscuring techniques make it more difficult to trace the trails of cryptocurrencies involved in illicit activities. For instance, if a large amount of money is stolen in a theft case, the stolen money will usually be split and sent to multiple addresses (accounts) and finally going back to the criminal's wallet. To investigate how money flows across different transactions or different blockchain platforms, researchers develop several cryptocurrency transaction tracing techniques.

The breadth-first search (BFS) algorithm and its variations have been applied in transaction tracing tasks. Zhao and Guan \cite{DBLP:conf/ifip11-9/ZhaoG15} use BFS to determine the most probable flowing directions of the stolen bitcoins in the Mt. Gox attack incident that happened in February 2014. 
Phetsouvanh et al. \cite{DBLP:conf/icdm/PhetsouvanhOD18} proposed an algorithm based on a variation of BFS for tracking flow confluences among some Bitcoin addresses. They applied this algorithm to the analysis of the 2015 Ashley Madison extortion scam, and further estimated the amount of money controlled by suspicious addresses in this scam.   
Haslhofer et al. \cite{DBLP:conf/i-semantics/HaslhoferKF16} proposed a Bitcoin transaction data analysis framework named GraphSense, which can trace the flows of bitcoins and find the connecting paths between given addresses or tags.

Transaction tracing techniques have also been proposed for taint analysis, whose principal aim is to predict the flow directions of the ``dirty money", which refers to the funds involved in illegal activities. If an address transfers the ``dirty money", it is considered to be tainted and should be blacklisted. The past researches proposed five methods for Bitcoin transaction tracing and tainting classification, namely the Poison method, Haircut method, First In First Out (FIFO) method \cite{DBLP:conf/fc/MoserBB14}, Last In First Out (LIFO) method, and Taint In Highest Out (TIHO) method \cite{DBLP:journals/corr/abs-1906-05754}. The description and evaluation of these methods are shown in Table \ref{TaintAnalysis}.

Figs. \ref{poison}-(e) show the examples for these five taint analysis methods. Each figure contains a transaction and presents the {\color{black}possible money flow directions between the inputs and outputs in the corresponding taint analysis strategy}. For example, according to the Poison method, all the outputs are tainted since the transaction has a tainted input in Fig. \ref{poison}, while for the haircut method, the amount value of tainted bitcoins stays the same in inputs and outputs. In Fig. \ref{FIFO}, the tainted input completely flows to the first output according to the chronological order. In a word, these methods can be seen as some prediction approaches aiming at tracing and making the ``dirty money" useless via blacklisting the possessors, so that the incentives for illicit activities within blockchain systems will be decreased. Tironsakkul et al. \cite{DBLP:journals/corr/abs-1906-05754} compared these five taint analysis methods and pointed out that these methods can be improved by incorporating other techniques like address clustering. Moreover, many other factors such as the distance from the original tainted address can be considered to determine the taint score \cite{DBLP:journals/corr/abs-1907-01538}.

Besides, Reid et al. \cite{Reid2013} proposed to trace the Bitcoin flows of an alleged theft by following the significant flows of bitcoins starting from some certain addresses. 
Meiklejohn et al. \cite{meiklejohn2013fistful} proposed an entity recognition algorithm which utilizes the change addresses to de-anonymize the money flows in Bitcoin, and traced the illicitly-obtained money of Silk Road and thefts by visualization analysis. 
With the help of automated trading platforms, money in blockchain systems can be transferred across different ledges. Yousaf et al. \cite{DBLP:conf/uss/YousafKM19} identified the cross-ledger money flows by finding out the transactions which happen close in time and have similar amount value. In particular, they summarized three different cross-ledger transaction patterns and proposed recognition methods for these cross-ledger transaction behaviors. 

{\color{black}Although some methods for transaction tracing have been mentioned, these methods are basically simple and heuristic. Moreover, all these methods except the method for cross-ledger transaction tracing \cite{DBLP:conf/uss/YousafKM19} are only suitable for Bitcoin-like blockchain platforms based on a transaction-centered model. Therefore, in the future, new transaction tracing methods based on risk propagation need to be designed under the consideration of more complex factors and scenarios.}
\section{Discussion \& Future Research Directions} \label{FRD}
\begin{table*}
	\renewcommand{\arraystretch}{1.5}
	\caption{\label{Summary} \color{black}Summary of cryptocurrency transaction network analysis methodologies in Section \ref{NM}-\ref{NBD}.}	
	\centering	
	\setlength{\tabcolsep}{0.9mm}{	
		\begin{tabular}{|p{0.27\columnwidth}|p{0.3\columnwidth}|p{0.45\columnwidth}|p{0.45\columnwidth}|p{0.48\columnwidth}|}
			\hline		
			\multicolumn{1}{|c|}{Category}& \multicolumn{1}{c|}{Subcategory} & \multicolumn{1}{c|}{Objective} & \multicolumn{1}{c|}{Main methodology} & \multicolumn{1}{c|}{Characteristic}\\		
			\hline
			\hline
			Network modeling &\multirow{3}*{\hspace{3.8em}-}&Modeling cryptocurrency transaction data as networks for downstream analysis tasks.&Approaches in Section \ref{NM}&Varying among different cryptocurrencies and lacking benchmarks\\ 
			\hline
			\multirow{10}*{Network profiling} &Network property analysis &Characterizing cryptocurrency transaction networks with static complex network measures.&Network measures in Section \ref{NPA}&\\ %
			\cline{2-4}
			&Network evolution analysis&Characterizing the dynamics of cryptocurrency transaction networks.&Uniform time sampling analysis in Section \ref{NEA}&Varying among different cryptocurrencies\\
			\cline{2-4}
			&Market effect analysis&Discussing the dynamic characteristics of the cryptocurrency market based on the transaction network.&Approaches in Section \ref{MEA}&\\
			\hline
			&Entity recognition&Clustering pseudonymous addresses sharing the same ownership.&Transaction property-based methods in Section \ref{ER} a)&Effective but easily to be attacked\\
			\cline{4-5}
			&&&Behavior-based methods in Section \ref{ER} b)&Coarse-grained address clustering\\
			\cline{4-5}
			&&&Off-chain information-based methods in Section \ref{ER} c)&Relaying on the off-chain information\\ %
			\cline{2-5}
			Network-based detection &Transaction pattern recognition&Recognizing the specific transaction behaviors.&Visualization methods in Section \ref{TPR} a)&Relying on visual observation\\
			\cline{4-5}
			&&&Tracking analysis in \ref{TPR} b)&Relying on tracking the behaviors of labeled addresses\\
			\cline{4-5}
			&&&Motif analysis in \ref{TPR} c)&Relying on pattern design and statistics on labeled addresses; High time complexity\\ 
			\cline{2-5}
			&Illicit activity detection&Detecting illicit activities and reporting the addresses involved in the activities.&Graph-based anomaly detection in Section \ref{IAD}&Relying on the analysis of specific illicit activities; Needing additional ways to handle the fast-growing data\\
			\cline{2-5}
			&Transaction tracing&Tracking the money flows across different transactions or different blockchain platforms.&Determining the probable flowing directions of money with methods such as BFS and taint analysis in Section \ref{TT}&Based on heuristic methods and easy to be attacked\\
			\hline
	\end{tabular}}
\end{table*}
{\color{black}Although a wealth of studies have been conducted, cryptocurrency transaction network analysis is still a challenging and promising research area. In this section, we discuss the above-mentioned studies focusing on their objectives and the characteristics of their methods. After that, we propose some possible future research directions based on the main challenges in this research issue.
\subsection{Summary and Discussion}
The research objective, main methodologies, and characteristics of the reviewed studies on cryptocurrency transaction network analysis are summarized in Table \ref{Summary}. 

As the foundation of cryptocurrency transaction network analysis, network modeling aims to abstract the cryptocurrency transaction data as a network by representing specific objects as nodes, and abstracting the relationships among the objects as edges. 
The transaction data of different cryptocurrencies are organized with distinct structures, and thus the abstracted networks vary a lot.
Although there are a variety of network modeling methods, an adaptive low-information loss modeling method suitable for most of the cryptocurrencies is lacked, which can provide a uniform input for downstream tasks.

Cryptocurrency transaction network profiling aims to extract descriptive information from the network and provide an overview of cryptocurrency trading. Existing studies under this topic can be mainly divided into network property analysis, network evolution analysis, and market effect analysis. These three aspects study cryptocurrency transaction networks by investigating their static network properties, dynamic evolution attributes, and their effects on the financial market. However, existing studies on network profiling are cryptocurrency-oriented and not comprehensive enough for altcoins since different cryptocurrencies have their own transaction network properties and market trend. Besides, blockchain techniques and cryptocurrency transaction data grow rapidly, which can result in the change of the properties in the original network. In particular, there is little discussion about the recent cryptocurrency transaction network after the emergence and successful development of 
DeFi, which has seriously affected the shape of the original cryptocurrency market as well as cryptocurrency transaction networks. 

Network-based detection on cryptocurrency transaction networks mainly contains four detective tasks including entity recognition, transaction pattern recognition, illicit activity detection, and transaction tracing. As the basis of many other downstream detective tasks, entity recognition aims to cluster pseudonymous addresses that may belong to the same user into an entity. Existing entity recognition methods mainly utilize the transaction properties, transaction behaviors, and off-chain information. In practical applications, transaction property-based methods cooperating with off-chain information are most frequently used. Although transaction property-based methods are effective in most cases, they can be easily bypassed via privacy-enhancing techniques such as mixing services.
For transaction pattern recognition, its main methodologies are transaction network visualization, tracking analysis, and motif analysis. There are also some unique limitations among these methods such as the high time complexity in motif matching. Based on the observed operation modes of illicit activities, graph-based anomaly detection methods are usually used in network-based illicit activity detection. However, the fast-growing transaction data hinder the deployment of illicit activity detection algorithms in real-time blockchain systems. Thus online algorithms \cite{crammer2003ultraconservative,ijcai2020-636} are needed to be developed in this area. For transaction tracing techniques which can track money flows across different transactions or different blockchain platforms, heuristic methods are usually used to determine the probable flowing direction of money. They also suffer from the problem that algorithms can be easily attacked by privacy-enhancing techniques, which is the common obstacles in network-based detective tasks. 

Since networks are expressive in describing interacting systems, up to now, there have been many studies exploring the cryptocurrency transaction data via transaction network mining. According to these studies, after modeling the cryptocurrency transaction data into complex networks, the efforts of existing studies can be categorized into two main types: 1) Characterizing cryptocurrency transaction networks to understand the transactions and financial market of cryptocurrencies. 2) Conducting detective tasks on cryptocurrency transaction networks to provide an insight into the related user behaviors and illicit activities, and providing help for building a healthier blockchain ecosystem.

\subsection{Future Research Directions}
Blockchain technology has huge potential in reforming the mode of traditional industries. In recent years, we have witnessed the rapid development of blockchain techniques and the fast growth of blockchain data size. However, existing cryptocurrency transaction network analysis methodologies still fail to meet the requirements of some important practical applications due to the main challenges brought by the multi-source heterogeneous blockchain data structure, the massive and rapid-increasing data, the trend of privacy enhancement, etc. Hence, we propose some possible future research directions which are worthy of further investigation.
\begin{enumerate}[1.]
    
	\item \textit{\textbf{Compatible transaction network modeling.}}
	The transaction model, data type, and data structure of different blockchain systems vary a lot. For example, compared with Bitcoin's transaction-centered model, Ethereum uses an account-centered model and introduces contract accounts. Transaction relationships in Ethereum are money transfer, contract creation, and contract invocation. In addition, in some blockchain systems like EOS, a transaction can contain multiple actions. Thus the heterogeneity of data from multiple sources brings great challenges to the design of network analysis methods. Besides, the downstream tasks in cryptocurrency transaction network analysis are also diverse. How to model with multi-source heterogeneous blockchain data to build support for different tasks is an important research direction. A new compatible cryptocurrency transaction network modeling is needed for different tasks and different cryptocurrencies.

	\item \textit{\textbf{Practical network-based information complement.}}
	The original architecture of blockchain systems has undergone massive changes with recent years' innovation of blockchain technology. Especially, many solutions are proposed to solve the scalability issues in blockchain systems \cite{DBLP:journals/access/ZhouHZB20}, such as Segregated Witness (SegWit) \cite{lombrozo2015bip141} and sharding \cite{luu2016secure,kokoris2018omniledger}. In some off-chain solutions like Lightning Network \cite{poon2016bitcoin} adopted by Bitcoin, incomplete transaction records are stored in the blockchain system. For example, the use of Lightning Network each time only results in two transaction records for the open and close status of a channel. However, it ignores the records of any transactions within the channel. Hence, a cryptocurrency transaction network may actually have incomplete link information. According to the ``comic effect'' \cite{tan2016efficient} of link prediction, the reshaped network with the addition of links predicted by link prediction algorithms can restore partially missing network structures and emphasize the important parts like an exaggerated but characteristic comic of the original network. Thus in the future, practical network-based information complement is a potential research direction in assisting cryptocurrency transaction analysis.
	
	\item \textit{\textbf{Dynamic transaction network analysis and online learning.}}
	Cryptocurrency transaction networks are currently the largest real-world networks with publicly accessible network data. Up to now, the most famous blockchain systems Bitcoin and Ethereum have accumulated hundreds of millions of transaction records. At the same time, cryptocurrency transaction networks are fast-growing, with new nodes and edges appearing constantly. The massive and rapid-increasing data bring great challenges to cryptocurrency transaction network analysis. On the one hand, method designs of transaction network analysis have to be scalable to large-scale transaction data. On the other hand, the large number of new emerging addresses and interactions can infect the properties of the original modeled network, and thus the analysis algorithms have to be updated adaptively according to the topology and attributes of the recent transaction network. Therefore, online algorithms \cite{crammer2003ultraconservative,ijcai2020-636} and scalable learning methods are needed to be developed in this research area in the future. Besides, the rich temporal information and the cold start problem can also be explored in future work on dynamic network analysis.
	
	\item \textit{\textbf{Feasible network-based transaction audit and tracing.}}
	The pseudonymous nature of blockchain systems prevents users from exposing their real identities in cryptocurrency transactions. In recent years, the growing need for privacy protection has also given birth to a lot of privacy preservation methodologies such as mixing services, ring signature \cite{rivest2001leak}, and non-interactive zero-knowledge proof \cite{feng2019survey}, making transactions more difficult to be traced. Although they allow blockchain users to achieve better privacy protection. On the other hand, the abuse of privacy-enhancing techniques can lead to rampant misbehaviors in blockchain systems, such as financial scams, money laundering, and so on. Transaction network analysis techniques have great potential in finding the relationship between objects and predicting the flow of money. Therefore, it is an important research direction to realize feasible transaction audit, illegal transaction tracing and interception with network-based methods in blockchain systems.
\end{enumerate}
}

\section{Conclusion} \label{C}
Since the debut of Bitcoin, cryptocurrency has been attracting increasing attention and wide acceptance worldwide. Thanks to the transparency and openness of blockchain, most of the cryptocurrency transactions are traceable and publicly accessible. By abstracting objects in the cryptocurrency system such as accounts, smart contracts, and entities as nodes, and the transaction behaviors between them as links, the cryptocurrency transactions among blockchain accounts are modeled as complex and large-scale transaction networks. In the past decade, academia has produced a large number of studies regarding cryptocurrency transactions from a network perspective. In this paper, we introduce the key concepts in cryptocurrency transactions, present a comprehensive review of the state-of-the-art literature on understanding the cryptocurrency transaction networks, and categorize the existing techniques and results into three closely related and mutually supportive steps in network analysis, i.e., network modeling, network profiling, and network-based detection. 

Despite the inspiring findings of the research questions in the existing literature, analysis, and mining of cryptocurrency transaction networks can also advance the development of complex network theory and graph mining techniques. To the best of our knowledge, cryptocurrency transaction networks are currently the largest real-world networks that can be built from publicly accessible data. Moreover, this kind of network has rapidly evolving network structures, abundant temporal, value and label information, providing network researchers with unprecedented opportunities for developing graph mining methodologies in this research area.  

With the rapid development of the cryptocurrency market and the scale of transaction data, we can foresee that more research attention will be devoted to studying cryptocurrency transactions. At the same time, large-scale graph mining technology is also a rapidly developing research field in recent years. As interdisciplinary research related to the above two fields, we believe that the study of cryptocurrency transaction networks is a promising area where an abundance of new findings, novel methods, and disruptive innovations will be seen in the future. Hopefully, this paper will serve as a reference and give researchers a systematical understanding of the key concepts and fundamental steps in cryptocurrency network analysis, thus become a well starting point to study in this field. 
 
\section*{Declaration of Interests}
The authors declare that they have no known competing financial interests or personal relationships that could have appeared to influence the work reported in this paper.


\bibliographystyle{cas-model2-names}

\bibliography{refs}

\end{document}